\documentclass[10pt, twocolumn, twoside, english]{IEEEtran}
\makeatletter
\def\normalsize{\@setfontsize{\normalsize}{9.7bp}{12.00pt}}
\normalsize
\makeatother

\usepackage{psfrag, epsfig}
\usepackage{array}
\usepackage{epsfig}
\usepackage{multicol}
\usepackage{multirow}
\usepackage{amsmath}
\usepackage{amssymb}
\usepackage{threeparttable}
\usepackage{bigstrut}
\usepackage{amssymb}
\usepackage[nolist , nohyperlinks]{acronym}
\usepackage{graphicx}
\usepackage{subfigure}
\usepackage{makecell}
\usepackage{siunitx}
\usepackage{hyperref}

\usepackage{algpseudocode}
\usepackage{algorithm}

\newcommand{\ie}{i.e.}
\newcommand{\eg}{e.g.}
\newcommand{\fig}{Fig.\,}
\newcommand{\scn}{Section\,}
\newcommand{\tbl}{Table\,}

\newcommand{\etal}{\emph{et al.\,}}

\graphicspath{{../figure/}}

\renewcommand{\baselinestretch}{.97}
\algnewcommand{\LineComment}[1]{\State \(\triangleright\) #1}

\usepackage{color}
\begin{document}
\title{LuMaMi28: Real-Time Millimeter-Wave Massive MIMO Systems with Antenna Selection}
\author{MinKeun~Chung,~\IEEEmembership{Member,~IEEE,}
        Liang~Liu,~\IEEEmembership{Member,~IEEE,}
        Andreas~Johansson,\\
        Sara~Gunnarsson,~\IEEEmembership{Student Member,~IEEE,}
        Martin~Nilsson,
        Zhinong~Ying,~\IEEEmembership{Fellow,~IEEE,}
        Olof~Zander,\\
        Kamal~Samanta~\IEEEmembership{Senior~Member,~IEEE},
        Chris~Clifton,
        Toshiyuki Koimori,
        Shinya Morita,
        Satoshi Taniguchi,\\
        Fredrik Tufvesson,~\IEEEmembership{Fellow,~IEEE,}
        and~Ove~Edfors,~\IEEEmembership{Senior~Member,~IEEE}
\thanks{M. Chung, L. Liu, A. Johansson, S. Gunnarsson, M. Nilsson, F. Tufvesson, and O. Edfors are with the Department of Electrical and Information Technology, Lund University, Lund, 221 00, Sweden (e-mail: \{firstname.lastname\}@eit.lth.se).}
\thanks{Z. Ying and O. Zander are with Sony Research Center Lund, Sweden (e-mail: \{firstname.lastname\}@sony.com).}
\thanks{K. Samanta and C. Clifton are with Sony Semiconductor and Electronic Solutions, Hampshire, UK (e-mail: \{firstname.lastname\}@sony.com).}
\thanks{T. Koimori, S. Morita, S. Taniguchi are with Sony Semiconductor Solutions, Analog LSI Business Division, Atsugi, Kanagawa, Japan (e-mail: \{firstname.lastname\}@sony.com).}
\thanks{This work is carried out within the Strategic Innovation Program “Smartare Elektroniksystem”, a joint venture of Vinnova, Formas and the Swedish Energy Agency (2018-01534).}
}

\maketitle

\begin{abstract}
This paper presents LuMaMi28, a real-time  \SI{28}{\GHz} massive multiple-input multiple-output (MIMO) testbed. In this testbed, the base station has 16 transceiver chains with a fully-digital beamforming architecture (with different pre-coding algorithms) and simultaneously supports multiple user equipments (UEs) with spatial multiplexing. The UEs are equipped with a beam-switchable antenna array for real-time antenna selection where the one with the highest channel magnitude, out of four pre-defined beams, is selected. For the beam-switchable antenna array, we consider two kinds of UE antennas, with different beam-width and different peak-gain. Based on this testbed, we provide measurement results for millimeter-wave~(mmWave) massive MIMO performance in different real-life scenarios with static and mobile UEs. We explore the potential benefit of the mmWave massive MIMO systems with antenna selection based on measured channel data, and discuss the performance results through real-time measurements.  
\end{abstract}

\IEEEpeerreviewmaketitle
\section{Introduction}
\label{sec:intro}
\ac{mmWave} communications, operating in the \SI{30}{\GHz}-\SI{300}{\GHz}, provide a promising approach. Since large bandwidths can be exploited, it achieves very high peak and cell-edge rates~\cite{singh15, ghosh15}. However, there are fundamental differences between \ac{mmWave} and traditional sub-\SI{6}{\GHz} communication systems. For example, propagation characteristics and hardware constraints. As it is widely known, \ac{mmWave} signals suffer from high \ac{FSPL}, which is inversely proportional to the wavelength $\lambda$~\cite{rap13}. The insertion losses and intrinsic power-overhead of \ac{mmWave} \acp{IC} result in diminishing gains~\cite{emil19}. Furthermore, the \ac{mmWave} system is more sensitive to hardware impairments, such as phase noise, \ac{PA} nonlinearities~\cite{Mchung2019_1, minkeun20_7, minkeun20_10}. It gives rise to technical challenges in the design and implementation of \ac{mmWave} systems. To address the challenges, we present LuMaMi28, a real-time \ac{mmWave} massive \ac{MIMO} systems that both analog- and digital-domain architectures are integrated.

\subsection{Related Work}
Over the past few years, the indoor\hspace{0.2em}/\hspace{0.2em}outdoor \ac{mmWave} channels have been extensively studied from propagation measurements~\cite{rap12, samimi16, rap17}. To overcome the high \ac{FSPL} and establish the links with sufficient \ac{SNR}, a large-scale antenna system, so-called massive \ac{MIMO}, should be needed for beamforming gain~\cite{swind14}. A smaller wavelength allows to use more antennas in the same physical space, resulting in higher array gains, as compared to the system with a larger wavelength. The massive \ac{MIMO} approach in \ac{mmWave} frequencies, however, is still prohibitive due to the high hardware cost and power consumptions~\cite{el14, choudhury15}. As a feasible solution to the problems, a hybrid beamforming architecture has been considered, which has a much lower number of \ac{TRx} chains than the total antenna number. The number of \ac{TRx} chains is only lower-limited by the number of data streams transmitted, while the beamforming gain is given by the number of antenna elements if suitable \ac{RF}\hspace{0.2em}/\hspace{0.2em}analog beamforming is done~\cite{molisch17}. Motivated by this fact, various hybrid beamforming architectures have been proposed in different papers~\cite{brady13, alkhateeb14, amadori15, venka16, sohrabi17}, to achieves near-optimal performance with lower cost and power consumption. Furthermore, finding an appropriate \ac{RF}\hspace{0.2em}/\hspace{0.2em}analog beamforming solution has been an active research area in recent years. \cite{vilar14, kwon16, hong17} have presented \ac{mmWave} antennas, and \cite{shinjo17, kamal_1, kamal_2} designed front-end modules, to support \ac{RF}\hspace{0.2em}/\hspace{0.2em}analog beamforming. With the \ac{RF}\hspace{0.2em}/\hspace{0.2em}analog beamforming strategies, \ac{mmWave}-specific baseband processing algorithms, such as channel estimation~\cite{alkhateeb14_2, vlachos19} and precoder~\cite{el14, alkhateeb14, gao16}, have been studied for cost- and energy-efficient systems. 
 
 A full implementation of \ac{mmWave} massive \ac{MIMO} systems, which includes \ac{RF}\hspace{0.2em}/\hspace{0.2em}analog- and digital-domain solutions, is an essential work for its practical use, but another significant challenge. Both industry and academia have been making efforts in building \ac{mmWave} massive \ac{MIMO} testbed. Samsung~\cite{roh14} and Qualcomm~\cite{raghavan18} have presented \SI{28}{\GHz} massive \ac{MIMO} prototypes, and provided their measurements and experimental results. Yang \etal~\cite{yang18} and Kuai \etal~\cite{kuai20} have presented a 64-\ac{TRx} fully-digital beamforming-based testbed, operating at \SI{28}{\GHz} and $37-42.5\hspace{0.1em}\rm{GHz}$ band, respectively. Recently, our research team has demonstrated the initial version of LuMaMi28 at IEEE Wireless Communication and Networking Conference in 2020~\cite{minkeun20_4}, and presented a \SI{28}{\GHz} hybrid beamforming testbed with a 64-antenna\hspace{0.1em}/\hspace{0.1em}16-\ac{TRx} unit~\cite{minkeun20_11}. However, to the best knowledge of the authors, none of the existing work on \ac{mmWave} massive \ac{MIMO} testbeds considers an \ac{UE}-side beam-steering approach.
\begin{figure}[t!]
\centering
\includegraphics[width = 3.54in]{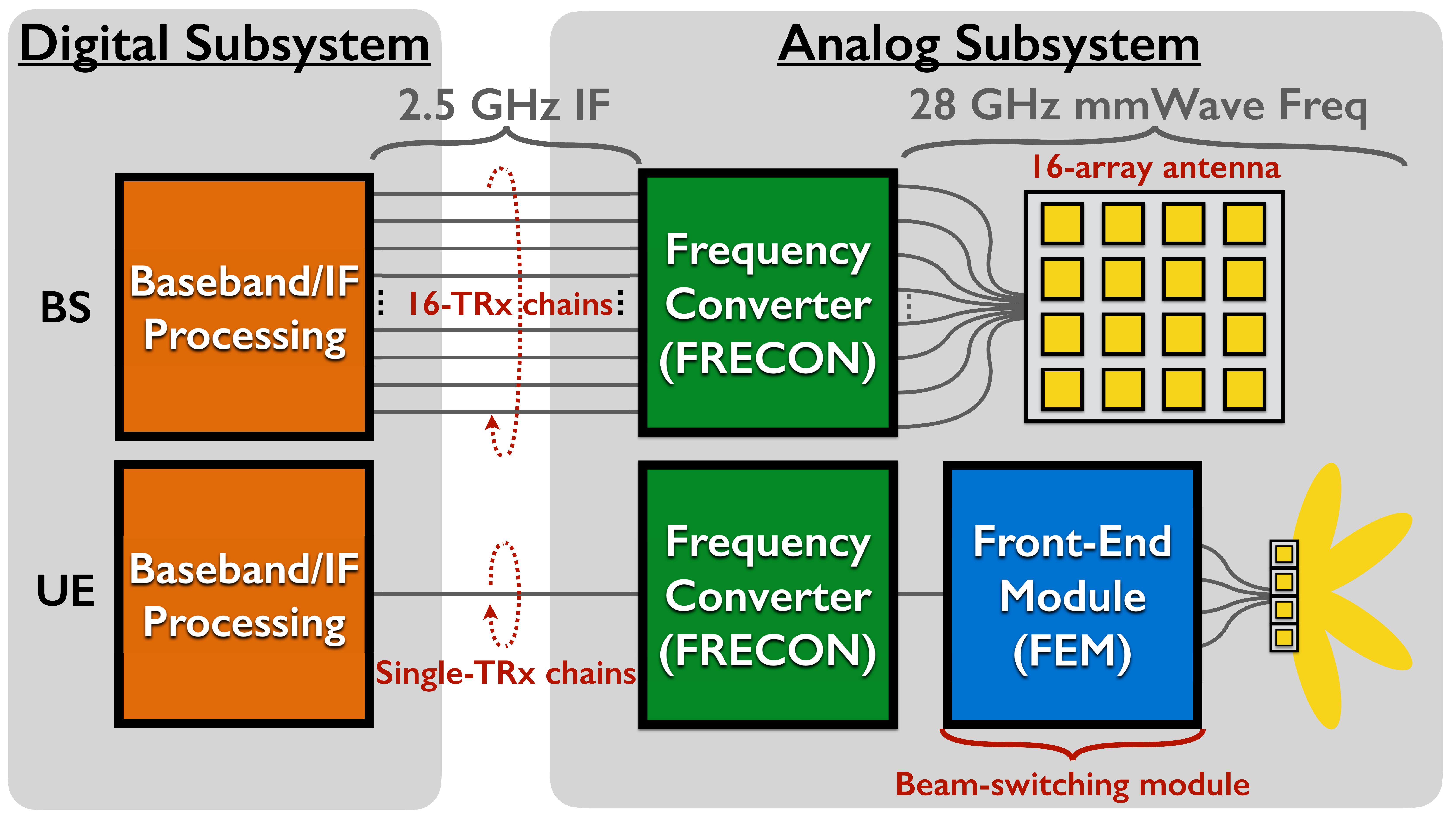}
\caption{System overview of our proposed mmWave massive MIMO testbed. A \ac{BS} and an \ac{UE}, respectively, consists of analog and digital subsystems.}
\label{fig:sys_testbed}
\vspace*{-0.25 cm}
\end{figure} 

\subsection{Contributions}
We develop LuMaMi28, a real-time \SI{28}{\GHz} massive \ac{MIMO} testbed, simultaneously supporting multiple \acp{UE}. \fig\ref{fig:sys_testbed} illustrates a high-level overview of LuMaMi28 \ac{BS} and \ac{UE}. The \ac{BS} has 16 \ac{TRx} chains with a fully-digital beamforming architecture. The \acp{UE} are equipped with a beam-switchable antenna array for antenna selection. To investigate the impact of different \ac{UE} antennas, we adopt yagi and patch \ac{UE} antennas, which have different beam-width and different peak-gain. Our main contributions are summarized as follows:
\begin{itemize}
	\item LuMaMi28 provides a flexible \ac{mmWave} massive \ac{MIMO} system for testing everything from baseband processing algorithms to scheduling and antenna selection in real-propagation environments. We extended the capability of an existing 100-antenna\hspace{0.1em}/\hspace{0.1em}100-\ac{TRx} massive \ac{MIMO} testbed~\cite{mal17}, operating below \SI{6}{\GHz}, to build LuMaMi28 \ac{BS}. The digital subsystem contains different equalizers\hspace{0.1em}/\hspace{0.1em}precoders, and the analog subsystem is designed in a modular way, for expanding to larger array sizes.
	\item LuMaMi28 \ac{UE} has a real-time beam-steering capability for antenna selection. The paper presents a co-design of analog\hspace{0.1em}/\hspace{0.1em}digital subsystems and its implementation, for this functionality. The proposed algorithm in the digital domain is designed on the \ac{FPGA} embedded in the \ac{SDR}, and the analog modules are developed in-house.
	\item We explore potential benefits achieved by the proposed antenna selection algorithm. To better assess the potential gains, a real-channel capture experiment was conducted, where the captured data is \ac{UL} channels between LuMaMi28 \ac{BS} and \ac{UE}.
	\item This paper presents real-time measurement results of LuMaMi28 in static and mobility environments. It studies the system performance with antenna selection, according to different digital-beamforming algorithms and different \ac{UE} antennas.
\end{itemize}

\subsection{Notation}
The set of complex numbers is denoted by $\mathbb{C}$. Lowercase boldface letters stand for column vectors and uppercase boldface letters designate matrices. For a vector or a matrix, we denote its transpose, conjugate, and conjugate transpose $\mathbf{(\cdot)}^{\rm T}$, $\mathbf{(\cdot)}^{*}$, and $\mathbf{(\cdot)}^{\rm H}$, respectively; The $N \times N$ identity matrix is denoted by $\mathbf{I}_{N}$. Sets are designated by upper-case calligraphic letter. A zero-mean circularly-symmetric complex-valued Gaussian random variable $x$, with variance $\sigma^{2}$, is denoted as $x\sim\mathcal{CN}(0, \sigma^{2})$.

\subsection{Outline of the Paper}
In the remainder of the paper is organized as follows. In Section \ref{sec:lumami28_ov}, we overview the architecture of LuMaMi28. Section \ref{sec:digital_sub}  and Section \ref{sec:analog_sub} describe the design and implementation of digital and analog subsystem, respectively, in LuMaMi28. In Section \ref{sec:measure}, we provide real-time measurement results of LuMaMi28, in different static and mobility scenarios. A summary and concluding remarks appear in Section \ref{sec:con}.

\section{LuMaMi28 Overview}
\label{sec:lumami28_ov}
This section provides an overview of LuMaMi28 architecture. LuMaMi28 consists of analog and digital subsystems, as shown in \fig\ref{fig:sys_testbed}. We employ the physical hardware setup of below-\SI{6}{\GHz} massive \ac{MIMO} testbed~\cite{mal17} to construct the digital subsystem including 16-\ac{TRx} chains. The analog subsystem developed in-house, is combined for supporting \ac{mmWave} bands. While the \ac{BS} has the fully-digital beamforming architecture, the \acp{UE} adopt a hybrid beamforming architecture for antenna selection, \ie, higher number of antennas than that of \ac{TRx} chain. Before delving into details, we refer the reader to \tbl\ref{table:sys_para}, which includes high-level system parameters of LuMaMi28.
\begin{table}[h]
\caption{High-level system parameters of LuMaMi28}
\renewcommand{\arraystretch}{1.4} 
\centering 
\setlength{\abovecaptionskip}{0pt}
\scalebox{0.9}{
\begin{threeparttable}
\begin{tabular} { >{\centering\arraybackslash}p{4.0cm} 
    >{\centering\arraybackslash}p{5.0cm}}
    \Xhline{3\arrayrulewidth}
    \bf{Parameter}& \bf{Value}\\
\hline
\centering{Carrier frequency} & \SI{27.95}{\GHz}\\ 
\centering{Intermediate frequency} & \SI{2.45}{\GHz}\\
\centering{Sampling frequency} & \SI{30.72}{\MHz}\\
\centering{Signal bandwidth} & \SI{20}{\MHz}\\
\centering{FFT size} & 2048\\
\centering{Antenna-array configuration} & 16 elements (\ac{BS})\hspace{0.3em}/\hspace{0.3em}4 elements (\ac{UE})\\
\centering{Number of \ac{TRx} chains} & 16 (\ac{BS})\hspace{0.33em}/\hspace{0.3em}1 (\ac{UE})\\
\centering{Number of maximum \acp{UE}} & 12\\
\centering{Power gain of FRECON} & $9\hspace{0.2em}\rm{dB}$~(Tx)\hspace{0.3em}/\hspace{0.3em}$7\hspace{0.2em}\rm{dB}$~(Rx)\\
\centering{Power gain of FEM} & $14\hspace{0.2em}\rm{dB}$~(Tx)\hspace{0.3em}/\hspace{0.3em}$12\hspace{0.2em}\rm{dB}$~(Rx)\\
\centering{Peak gain of a \ac{BS} antenna} & $5\hspace{0.2em}\rm{dBi}$\\
\centering{Peak gain of \ac{UE}1\hspace{0.3em}/\hspace{0.3em}\ac{UE}2 antennas} & $7.5\hspace{0.2em}\rm{dBi}$ (yagi)\hspace{0.33em}/\hspace{0.3em}$10.0\hspace{0.2em}\rm{dBi}$ (patch)\\ 
\centering{Power gain of each Tx chain} & $22\hspace{0.2em}\rm{dB}$ (linear region)\\ 
\centering{P1dB of each Tx chain} & $18\hspace{0.2em}\rm{dBm}$\\ 
\centering{Frame time} & \SI{10}{\ms}\\
\centering{Beam sweeping duration} & \SI{10}{\ms} (\SI{40}{\ms} for four \ac{RF} ports)\\ 
\centering{Guard time for beam switching} & \SI{71.9}{\us}\\  
\Xhline{3\arrayrulewidth} 
\end{tabular}
\end{threeparttable}
}
\vspace*{-.2cm}
\label{table:sys_para}
\end{table}
\begin{figure*}[t!]
\centering
\includegraphics[width = 6.45in]{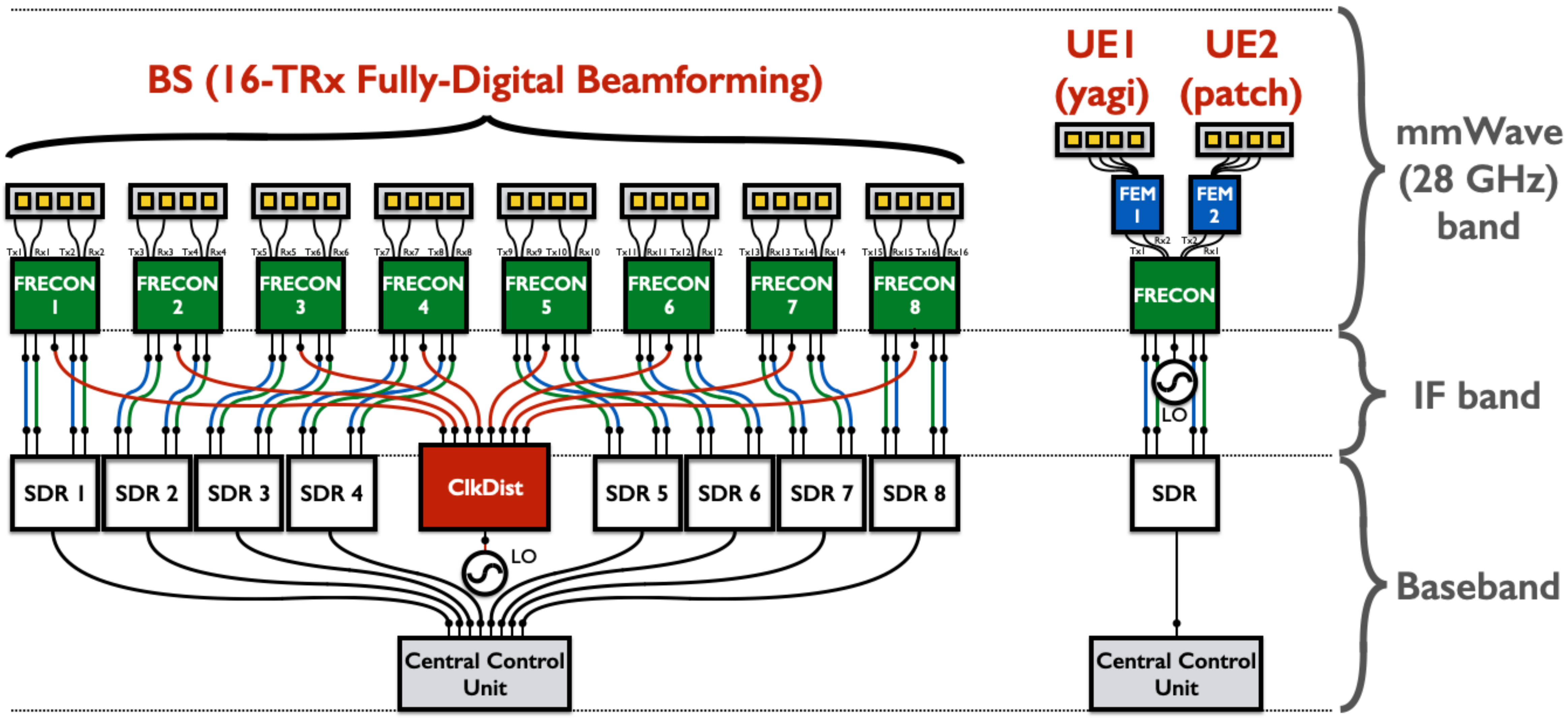}
\caption{Architectures of LuMaMi28 \ac{BS}~(16-\ac{TRx} fully-digital beamforming) and \acp{UE}~(antenna selection by using a 4-antenna\hspace{0.1em}/\hspace{0.1em}1-\ac{TRx} unit). This setup were used for measurement campaigns that we will present in \scn\ref{sec:measure}. DC control signals from each SDR, \ie, for TDD and beam switching, are delivered to FRECON~(only TDD switching signal) and FEM (both). For simplicity, the routes for the DC control signals are omitted in this figure.}
\label{fig:Arch_testbed}
\vspace*{-0.25 cm}
\end{figure*}

The digital subsystem is in charge of baseband and \ac{IF} processing. A central control unit has an embedded controller (NI PXIe-8135) or a laptop, for the \ac{BS} and \ac{UE}, respectively, which runs \emph{LabVIEW} on a standard Windows 7 64-bit operating system to configure and control LuMaMi28. LabVIEW provides both host and \ac{FPGA} programming. The central control unit for the \ac{BS} and \ac{UE}, respectively, is plugged into eight and one \acp{SDR}~(NI USRP-294xR\hspace{0.1em}/\hspace{0.1em}295xR). Each \ac{SDR} is equipped with two \ac{TRx} chains and a Kintex-7 \ac{FPGA}. The \ac{SDR} performs local processing on a per-\ac{TRx} basis, \eg, \ac{OFDM} processing and reciprocity calibration. In the central control unit of the \ac{BS}, co-processing \ac{FPGA} modules~(FlexRIO 7976R) is included for digital-beamforming processing. To be able to synchronize 16 \ac{TRx} chains, a reference clock source (PXIe-6674T) and reference clock distribution network~(Octo-Clock) are used.

The \ac{BS} and \ac{UE} analog subsystems contain in common identical \acp{FRECON} and a common \acp{LO}, respectively, for up/down conversion between \ac{IF} and \SI{28}{\GHz} bands. The \ac{BS} has a \ac{ClkDist} for amplifying and distributing  the \ac{LO} signal to multiple \acp{FRECON}, and each \ac{UE} has a \ac{FEM} for beam steering. For  reconfigurability and scalability of LuMaMi28, we designed the \ac{FRECON} and \ac{FEM} \acp{PCB} in a modular way. Each module is equipped with a small number of \ac{TRx} chains, \ie, two per \ac{FRECON} and one per \ac{FEM}. A basic configuration is that, therefore, one \ac{FRECON} is plugged with one \ac{SDR}, which can construct two \ac{TRx} chains for the \ac{BS}, and two \acp{UE} with single \ac{TRx} chain. Both \ac{FRECON} and \ac{FEM} have SPDT switches for \ac{TDD} operation while the \ac{FEM} has additionally a SP4T switch for switching analog-domain beams. The \ac{FRECON} contains up\hspace{0.1em}/\hspace{0.1em}down conversion mixers, \acp{LPF}, \acp{DA}, and \acp{LNA}; the \ac{FEM} a \ac{PA} , an \ac{LNA}, and an \ac{LPF} for achieving better output power and noise-floor performance. Eight \acp{FRECON} in the \ac{BS} plugs into a 16-element array antenna, and \ac{FEM} in each \ac{UE} into a 4-element array antenna. As mentioned earlier, yagi and patch antennas are used to explore the impact of different \ac{UE} antennas on system performances.

To control the analog subsystem, \ie, for \ac{TDD} switching and antenna selection, we adopt wired links between analog and digital subsystems. DC control signals~(3.3V) are operated by an implemented control unit in the digital subsystem. We use a 15-pin \ac{GPIO} port in each \ac{SDR} to deliver the control signals. \fig\ref{fig:Arch_testbed} illustrates the architecture of LuMaMi28 \ac{BS} and \ac{UE} that we used for measurements.

\section{Digital Subsystem Design and Implementation}
\label{sec:digital_sub}
In this section, the digital subsystem details in LuMaMi28 are discussed based on the aforementioned  testbed architecture. To build the digital subsystem, we need to make a few design decisions. First, we reused in part the frame structure and baseband functionalities, such as \ac{OFDM}, time/frequency synchronization and precoders/equalizers, of the below-\SI{6}{\GHz} massive \ac{MIMO} testbed~\cite{mal17}. Second, a LuMaMi28 \ac{UE} has the beam-switching capability in the analog subsystem. To operate this, we design an antenna selection algorithm and implement the proposed algorithm in the digital subsystem. Third, notice that control signals for \ac{TDD} and beam-switching need to be sent from the digital to the analog subsystem. Hence an analog-subsystem control unit has to be implemented in the digital subsystem. In the following subsection, we describe key precoders and equalizers built in the digital subsystem of LuMaMi28.

\subsection{Baseband Precoding and Equalization}
We consider a simplified model of a \ac{MU}-\ac{MIMO} system. The system includes one \ac{BS} equipped with $M$ \ac{TRx}-chains simultaneously serving $K$ single \ac{TRx}-chain \ac{UE} in \ac{TDD} operation. To reduce the channel-estimation overhead, we adopt a fully digital reciprocity-based beamforming. Based on the fact that an \ac{UL} propagation-channel matrix $\mathbf{H}_{\mathsf{p}} \in \mathbb{C}^{M \times K}$ is reciprocal, the estimated \ac{UL} channel can be exploited to compute precoding matrices for \ac{DL} transmission. The $K$ \acp{UE} transmit their \ac{UL} data $\mathbf{s}_{\mathsf{ul}} \in \mathbb{C}^{K \time 1}$ in the same time-frequency resource. The average \ac{UL} power levels used by the $K$ \acp{UE} during transmission are represented by $\mathbf{p}_{\mathsf{ul}} = [p_{1}, p_{2}, \cdots, p_{K} ]^{\rm T}$. The received signals $\mathbf{y}_{\mathsf{bs}} \in \mathbb{C}^{M \times 1}$ at the \ac{BS} is
\begin{align}
\label{eq:ul_rx_model}
\mathbf{y}_{\mathsf{bs}}
= 
\mathbf{H}_{\mathsf{ul}}
\sqrt{\mathbf{P}_{\mathsf{ul}}}
{\hspace{0.2em}}
\mathbf{s}_{\mathsf{ul}}
+
\mathbf{z}_{\mathsf{bs}},
\end{align}
where $\mathbf{H}_{\mathsf{ul}} \in \mathbb{C}^{M \times K}$ is the \ac{UL} channel{\footnote{$\mathbf{H}_{\mathsf{ul}}$ represents the \ac{UL} radio channel including both the propagation channel $\mathbf{H}_{\mathsf{p}}$ and the \ac{UL} hardware transfer functions. Its factorization is shown in (\ref{eq:ul_channel_fact}).}} $\mathbf{P}_{\mathsf{ul}} \in \mathbb{C}^{K \times K}$ the diagonal matrix with the entries of $\mathbf{p}_{\mathsf{ul}}$ on its main diagonal, $\mathbf{z}_{\mathsf{bs}} \in \mathbb{C}^{K \times 1}$ the \ac{AWGN} with i.i.d. $\mathcal{CN} (0,1)$ entries. The detected \ac{UL} symbols $\hat{\mathbf{s}}_{\mathsf{ul}} \in \mathbb{C}^{K \times 1}$ by using an equalization matrix $\mathbf{F}_{\mathsf{eq}} \in \mathbb{C}^{K \times M}$ is 
\begin{align}
\label{eq:detect_UL_sym}
\hat{\mathbf{s}}_{\mathsf{ul}}
= 
\mathbf{F}_{\mathsf{eq}}
\mathbf{y}_{\mathsf{bs}}.
\end{align}
The received signal vector at the \acp{UE} in the \ac{DL} $\mathbf{y}_{\mathsf{dl}} \in \mathbb{C}^{K \times 1}$ is
\begin{align}
\label{eq:dl_rx_model}
\mathbf{y}_{\mathsf{ue}}
= 
\mathbf{H}_{\mathsf{dl}}
\mathbf{F}_{\mathsf{pr}}
{\hspace{0.2em}}
\mathbf{s}_{\mathsf{dl}}
+
\mathbf{z}_{\mathsf{ue}},
\end{align}
where $\mathbf{H}_{\mathsf{dl}} \in \mathbb{C}^{K \times M}$ is the \ac{DL} channel, $\mathbf{F}_{\mathsf{pr}} \in \mathbb{C}^{M \times K}$ the precoding matrix at \ac{BS}, $\mathbf{s}_{\mathsf{dl}} \in \mathbb{C}^{K \times 1}$ the \ac{DL} data vector intended for the $K$ \acp{UE}, $\mathbf{z}_{\mathsf{ue}} \in \mathbb{C}^{K \times 1}$ the \ac{AWGN} with i.i.d. $\mathcal{CN} (0,1)$ entries. Note that the \ac{UL} and \ac{DL} channels are generally not reciprocal in practical systems due to differences of analog circuitry in \ac{UL} and \ac{DL} channels as follows.
\begin{align}
\label{eq:ul_channel_fact}
\mathbf{H}_{\mathsf{ul}}
= 
\mathbf{R}_{\mathsf{bs}}
\mathbf{H}_{\mathsf{p}}
\mathbf{T}_{\mathsf{ue}},
\end{align}
\begin{align}
\label{eq:dl_channel_fact}
\mathbf{H}_{\mathsf{dl}}
=
\mathbf{R}_{\mathsf{ue}}
\mathbf{H}_{\mathsf{p}}^{\rm T}
\mathbf{T}_{\mathsf{bs}}.
\end{align}
where the diagonal matrices $\mathbf{R}_{\mathsf{bs}} \in \mathbb{C}^{M \times M}$ and $\mathbf{R}_{\mathsf{ue}} \in \mathbb{C}^{K \times K}$ model the non-reciprocal hardware responses of \ac{BS} and \ac{UE} receivers, respectively; the diagonal matrices $\mathbf{T}_{\mathsf{bs}} \in \mathbb{C}^{M \times M}$ and  $\mathbf{T}_{\mathsf{ue}} \in \mathbb{C}^{K \times K}$ hardware responses of \ac{BS} and \ac{UE} transmitters, respectively. From (\ref{eq:ul_channel_fact}) and (\ref{eq:dl_channel_fact}), the precoding matrix includes a calibration matrix $\mathbf{C} \triangleq {\mathbf{T}_{\mathsf{bs}}} {\mathbf{R}_{\mathsf{bs}}^{-1}}$~\cite{vie17}. For the digital subsystem, we consider three linear equalizers\hspace{0.1em}/\hspace{0.1em}precoders, \ac{MRC}\hspace{0.1em}/\hspace{0.1em}\ac{MRT}, \ac{ZF}, and \ac{RZF}, \ie,
\begin{align}
\label{eq:linear_eq}
\mathbf{F}_{\mathsf{eq}} = 
\left\{
	\begin{array}{ll}
    	\displaystyle{\mathbf{H}_{\mathsf{ul}}^{\rm H}} & \text{for \ac{MRC} }\\
        \vspace{0.01cm}    &  \\
        \displaystyle{\big(\mathbf{H}_{\mathsf{ul}}^{\rm H}\mathbf{H}_{\mathsf{ul}}\big)^{-1} \mathbf{H}_{\mathsf{ul}}^{\rm H}} & \text{for \ac{ZF} }\\
        \vspace{0.01cm}    &  \\
        \displaystyle{\big(\mathbf{H}_{\mathsf{ul}}^{\rm H} \mathbf{H}_{\mathsf{ul}} + \alpha_{\mathsf{eq}} \mathbf{I}_{K}\big)^{-1} \mathbf{H}_{\mathsf{ul}}^{\rm H}} & \text{for \ac{RZF} }
    \end{array}
\right.,
\end{align}
\begin{align}
\label{eq:linear_pc}
\mathbf{F}_{\mathsf{pr}} = 
\left\{ 
	\begin{array}{ll}
    	\displaystyle{\mathbf{C} \mathbf{H}_{\mathsf{ul}}^{*}} & \text{for \ac{MRT} }\\
        \vspace{0.01cm}    &  \\
        \displaystyle{\mathbf{C} \mathbf{H}_{\mathsf{ul}}^{*} \big(\mathbf{H}_{\mathsf{ul}}^{\rm H}\mathbf{H}_{\mathsf{ul}}\big)^{\rm{-T}} } & \text{for \ac{ZF} }\\
        \vspace{0.01cm}    &  \\
        \displaystyle{\mathbf{C} \mathbf{H}_{\mathsf{ul}}^{*} \big(\mathbf{H}_{\mathsf{ul}}^{\rm H} \mathbf{H}_{\mathsf{ul}} + \alpha_{\mathsf{pr}} \mathbf{I}_{K}\big)^{\rm {-T}} } & \text{for \ac{RZF} }
    \end{array}
\right.,
\end{align}
where $\alpha_{\mathsf{eq}}$ and $\alpha_{\mathsf{pr}}$, respectively, is a regularization constant that allows trade-off between array gain and interference suppression. The optimal trade-off of $\alpha_{\mathsf{eq}}$ and $\alpha_{\mathsf{pr}}$ results in minimum \ac{MSE} precoder/equalizer~\cite{peel05}.

\begin{figure}[t!]
\subfigure[Generic structure]
{
\includegraphics[width=3.39in]{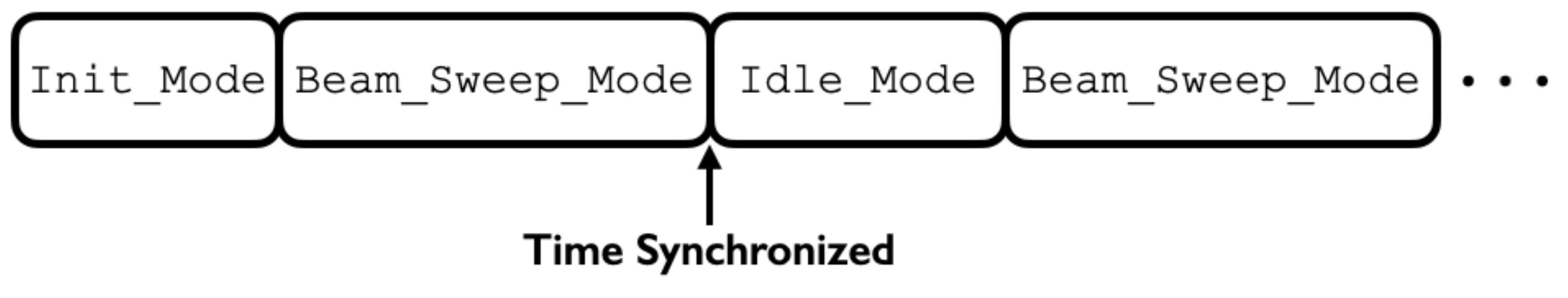}
\label{fig:gen_struct}
}
\vspace*{0.25 cm}
\centering
\subfigure[$\texttt{Init\_Mode}$]
{
\includegraphics[width=2.59in]{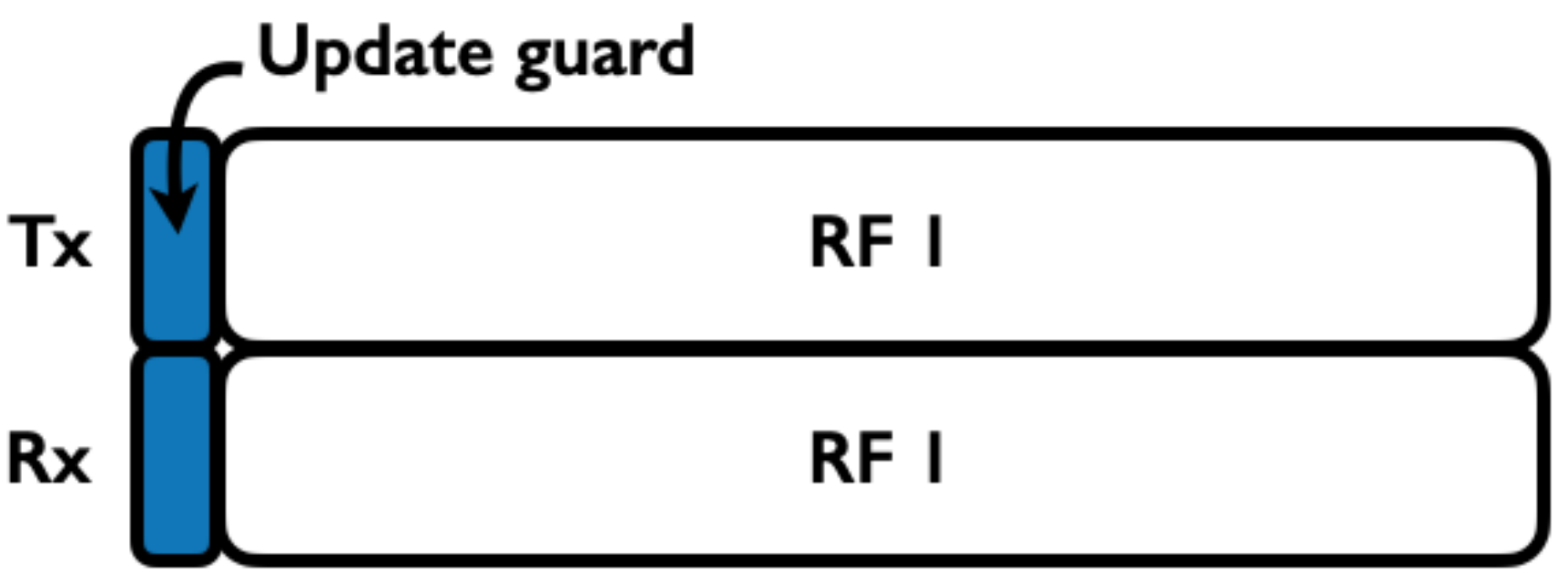}
\label{fig:init_mode}
}
\vspace*{0.25 cm}
\subfigure[$\texttt{Beam\_Sweep\_Mode}$]
{
\includegraphics[width=2.59in]{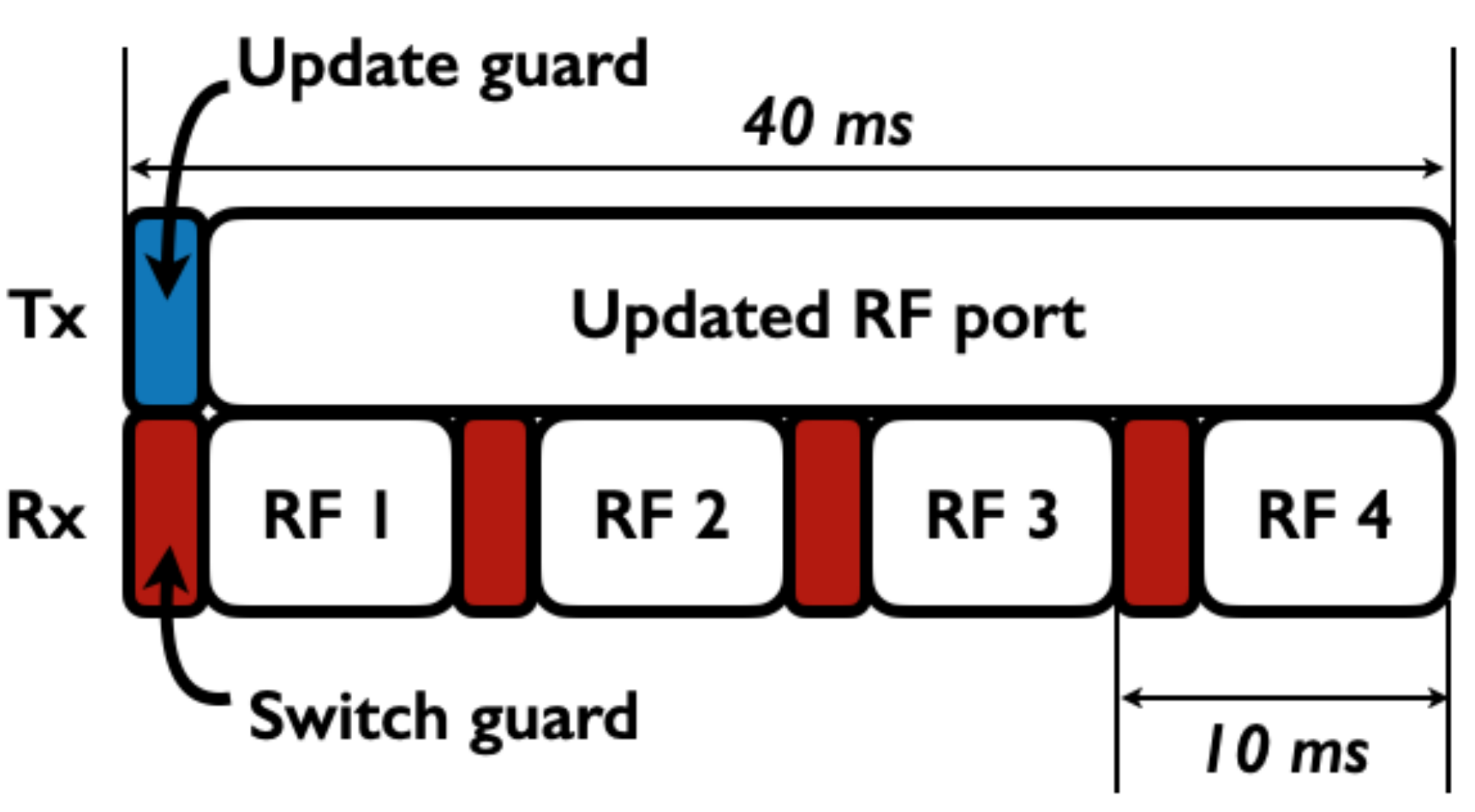}
\label{fig:beam_sweep_mode}
}
\vspace*{0.25 cm}
\subfigure[$\texttt{Idle\_Mode}$]
{
\includegraphics[width=2.59in]{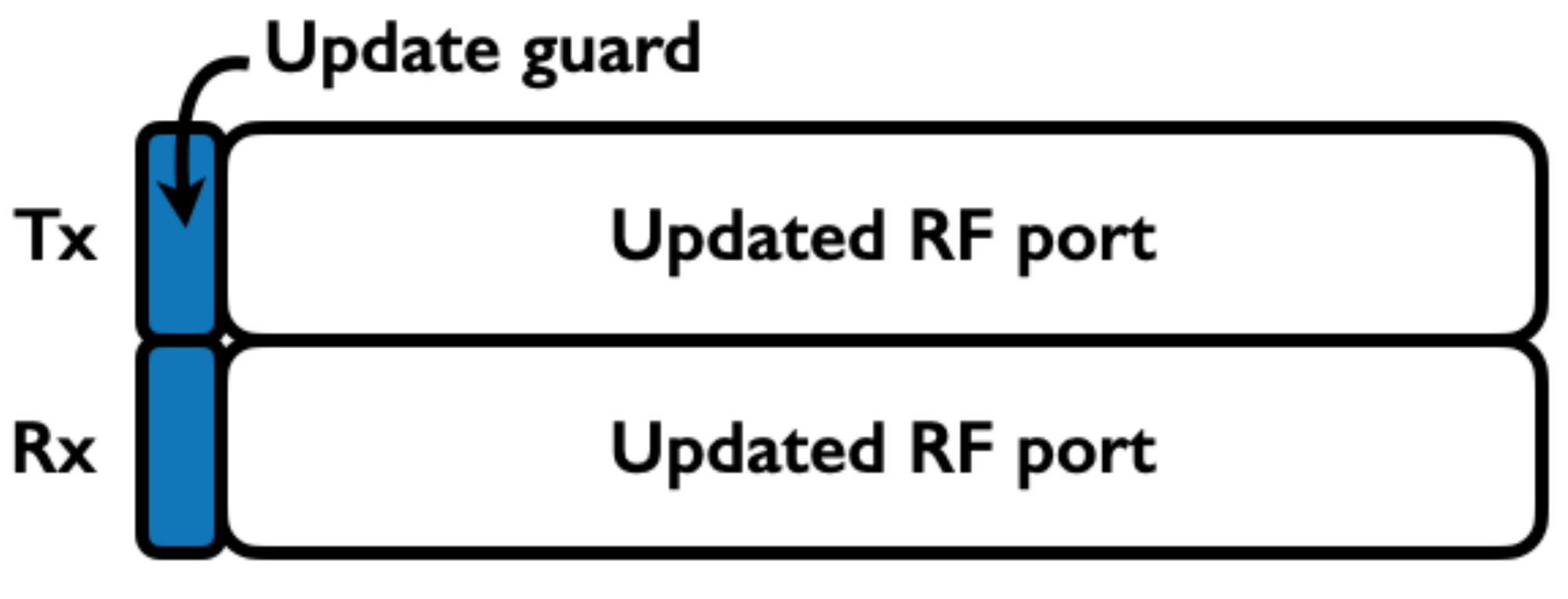}
\label{fig:idle_mode}
}
\caption{Antenna selection mode. Each mode has an update guard~(blue) or a switch guard time~(red) for beam sweeping or updating.} 
\label{fig:Ant_Select_Mode}
\vspace*{-0.25 cm}
\end{figure}

\subsection{Antenna Selection Algorithm}
The basic idea for antenna selection is as follows. Each \ac{UE} performs a regular beam sweeping in the reception mode. The \ac{DL} channel estimation block calculates the channel magnitude for each beam. Based on the calculation, the \ac{UE} selects the one with highest channel magnitude. For real-time operation, the antenna selection algorithm should ensure that the beam sweeping and its channel-magnitude calculation can be synchronized, further, a selected beam can be updated on time. This subsection describes an architecture of the antenna selection algorithm.

\begin{figure}[t!]
\centering
\includegraphics[width = 3.39in]{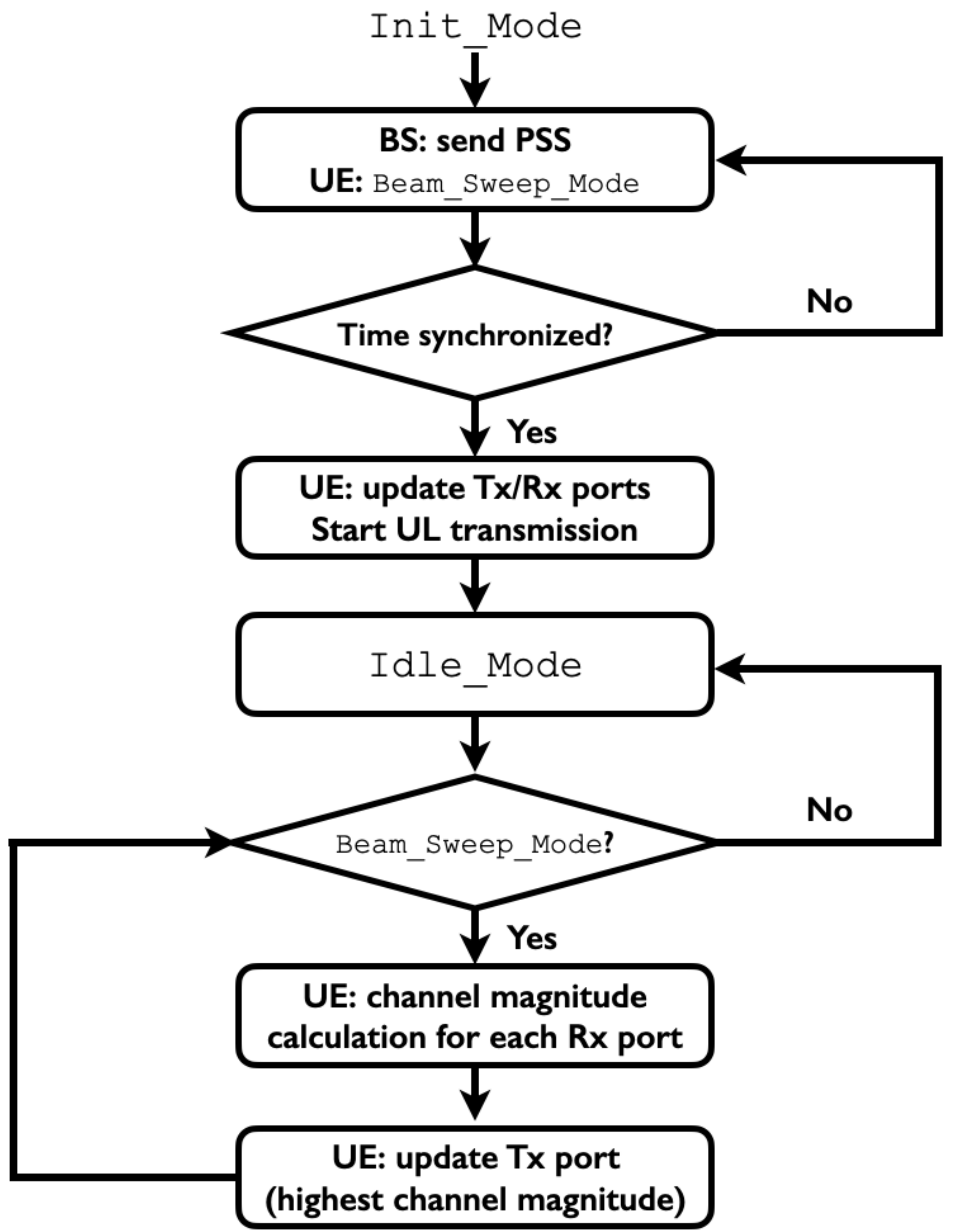}
\caption{Flow chart for our antenna selection implementation.}
\label{fig:fc_ant_select}
\vspace{-0.25 cm}
\end{figure}

The proposed algorithm is designed on the \ac{FPGA} embedded in the \ac{SDR} and integrated with baseband functionalities of each \ac{UE}. For its implementation, we define three modes regarding antenna selection process, as illustrated in \fig \ref{fig:Ant_Select_Mode}. All modes include an update guard or a switch guard time for beam updating and sweeping, respectively. Both the guard time slots have one \ac{OFDM} symbol duration (\SI{71.9}{\us}). First, $\texttt{Init\_Mode}$ is to set up an initial \ac{RF} port. We employ the first \ac{RF} port (\ac{RF}1) as the initial port for both \ac{Tx} and {Rx}. As mentioned in \scn \ref{sec:lumami28_ov}, the \ac{FEM} is switchable among four \ac{RF} ports. $\texttt{Beam\_Sweep\_Mode}$ is to switch among the four \ac{Rx} ports where each \ac{RF} port is switched every \SI{10}{\ms} (one frame{\footnote{As mentioned earlier, we reused the frame structure of the existing below-\SI{6}{\GHz} massive testbed~\cite{mal17}, which is \ac{LTE} like \ac{TDD}-based frame structure. We adopt one frame duration to synchronize beam-sweeping operation and its channel-magnitude calculation in an effective way.}) in order. Notice that there are two different targets in the use of $\texttt{Beam\_Sweep\_Mode}$. One is for aligning beams before timing synchronization. The other is for calculating \ac{DL} channel magnitudes for each beam, after timing synchronization. A narrow beam direction at both a \ac{mmWave} \ac{BS} and \acp{UE} results in low \ac{SNR} before beamforming, \ie, in isotropic propagation conditions~\cite{hur13}. Hence, a beam alignment through $\texttt{Beam\_Sweep\_Mode}$ is required for timing synchronization. In this $\texttt{Beam\_Sweep\_Mode}$, without channel-magnitude calculation, the index of \ac{RF} port used for a successful time synchronization is updated for both \ac{Tx} and \ac{Rx} mode. It is the difference with the other $\texttt{Beam\_Sweep\_Mode}$. During $\texttt{Beam\_Sweep\_Mode}$, the \ac{RF} port for transmission is determined by the result of previous beam switching, and its updated beam is kept (in case of $\texttt{Beam\_Sweep\_Mode}$ before time synchronization, initial \ac{RF} port, \ie, \ac{RF}1). Lastly, $\texttt{Idle\_Mode}$ does not include switching operation and its period is adjustable unlike $\texttt{Beam\_Sweep\_Mode}$ (\SI{40}{\ms}). Using an updated beam from the previous mode, the \ac{RF} port for both transmission and reception is fixed during $\texttt{Idle\_Mode}$. 

\fig \ref{fig:fc_ant_select} abstracts the architecture of the antenna selection algorithm. For \ac{OTA} synchronization of \ac{BS} and \acp{UE}, we exploit the \ac{LTE} \ac{PSS} generated from a frequency-domain Zadoff-chu sequence~\cite{sesia2009lte}. After $\texttt{Init\_Mode}$, the LuMaMi28 \ac{BS} broadcasts the \ac{PSS}, and \acp{UE} enter $\texttt{Beam\_Sweep\_Mode}$ for the beam alignment. If the start of the frame is estimated by the \ac{OTA} time synchronization, each \ac{UE} starts the \ac{UL} transmission, and operates in $\texttt{Idle\_Mode}$ until before the next $\texttt{Beam\_Sweep\_Mode}$ for calculating \ac{DL} channel magnitudes. The beam updated for this $\texttt{Idle\_Mode}$ is formed by the \ac{RF} port that makes a success of time synchronization. We design the $\texttt{Beam\_Sweep\_Mode}$, operated after being time-synchronized, so that the switch of \ac{RF} port, its channel-magnitude calculation, and the antenna-index update of the higher channel magnitude can be accommodated within one frame. Furthermore, all the above operations are synchronized by using a sample counter, resulting in preventing a mismatch between the operations. For example, incorrect mapping between a switched beam and its channel-magnitude calculation, and delayed update of selected beam.  

\begin{figure}[t!]
\centering
\includegraphics[width = 3.39in]{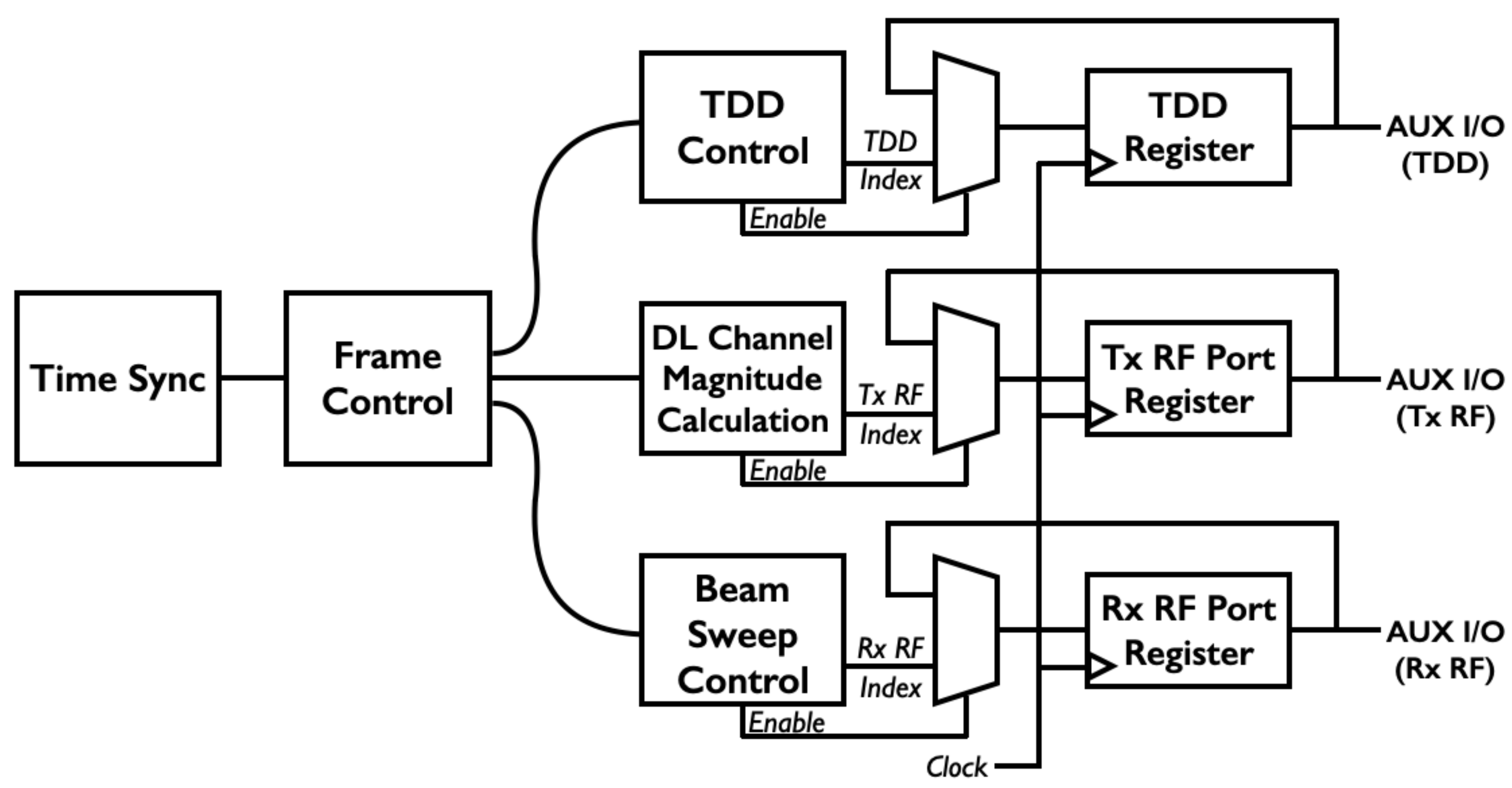}
\caption{Block diagram of analog-subsystem control unit. Since an 1-bit index for the \ac{TDD} switching, and a 2-bit index for \ac{RF} \ac{Tx} and \ac{Rx}, respectively, for the antenna selection, are needed, the same number of registers (five in total) are required to control the analog subsystem. It is simplified in this figure.}
\label{fig:bd_analog_control}
\vspace{-0.25 cm}
\end{figure}

\subsection{Analog-Subsystem Control Unit}
There is two control signals to be delivered from the digital to the analog subsystem. One is an 1-bit \ac{TDD} index, and the other a 2-bit \ac{RF}-index for the antenna selection. We also implemented an analog-subsystem control unit on the \ac{FPGA} embedded in the \ac{SDR}. \fig \ref{fig:bd_analog_control} illustrates a simplified block diagram of the implemented analog-subsystem control unit. The frame control unit includes the master sample-counter mentioned in the previous subsection, and play a role in synchronizing the following three units. In the DL channel magnitude calculation and beam sweep control units, there is a slave counter, respectively, for $\texttt{Beam\_Sweep\_Mode}$. Each index is written to the dedicated register with 2-to-1 multiplexer, followed by an AUX I/O connection. 

Both \ac{FRECON} and \ac{FEM} have an interface, respectively, which are connected with the \ac{GPIO} port. The interface in \ac{FRECON} is for receiving only \ac{TDD} control signal, and the one in \ac{FEM} for both \ac{TDD} switching and antenna selection. Since the \ac{GPIO} port plugs in to AUX I/O ports of an \ac{FPGA} embedded in each \ac{SDR}, the DC signal~(\SI{3.3}{\V}) is controllable according to the implemented analog-subsystem control unit in the digital subsystem. To formulate the delay of control signal from digital to analog subsystem, the following four delays introduced from hardware units needs to be considered: 1) delay on rising (falling) edge in the digital subsystem ($\Delta_{\mathsf{d}}$), 2) SPDT switching delay of \ac{FRECON} ($\Delta_{\mathsf{sd}}^{\mathsf{fre}}$), 3) SPDT switching delay of \ac{FEM} ($\Delta_{\mathsf{sd}}^{\mathsf{fem}}$), and 4) SP4T switching delay of \ac{FEM} ($\Delta_{\mathsf{s4}}^{\mathsf{fem}}$). The total delay from digital subsystem to \ac{FRECON} and \ac{FEM}, respectively, is
\begin{align}
\label{eq:frecon_delay}
\Delta_{\mathsf{frecon}}
= 
\Delta_{\mathsf{d}}
+
\Delta_{\mathsf{sd}}^{\mathsf{fre}},
\end{align}
\begin{align}
\label{eq:fem_delay}
\Delta_{\mathsf{fem}}
= 
\Delta_{\mathsf{d}}
+
\Delta_{\mathsf{sd}}^{\mathsf{fem}}
+
\Delta_{\mathsf{s4}}^{\mathsf{fem}}.
\end{align}
Based on the measurement, each delay results in $\Delta_{\mathsf{d}} \simeq \SI{100}{\ns}$, $\Delta_{\mathsf{sd}}^{\mathsf{fre}} \simeq \SI{20}{\ns}$, $\Delta_{\mathsf{sd}}^{\mathsf{fem}} \simeq \SI{85}{\ns}$, and $\Delta_{\mathsf{s4}}^{\mathsf{fem}} \simeq \SI{200}{\ns}$, respectively. It is shown that the guard time (\SI{71.9}{\us}) is enough for \ac{TDD} switching and antenna selection ($\Delta_{\mathsf{frecon}} \simeq \SI{120}{\ns}$ and $\Delta_{\mathsf{fem}} \simeq {\SI{385}{\ns}}$). 
\begin{figure}[t!]
\centering
\subfigure[ ]
{
\includegraphics[width=3.43in]{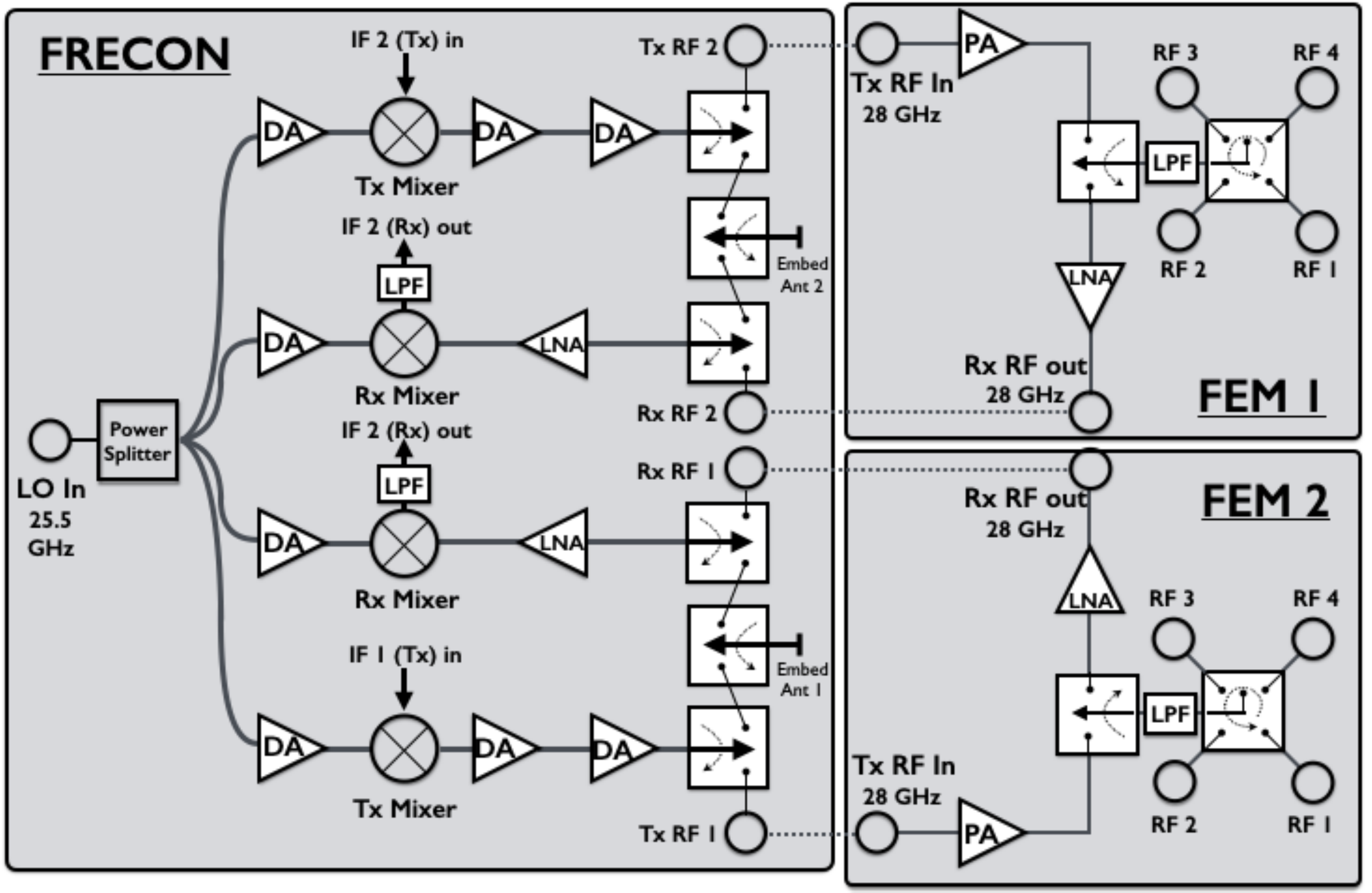}
\label{fig:BD_FRECON_FEM}
}
\subfigure[ ]
{
\includegraphics[width=3.43in]{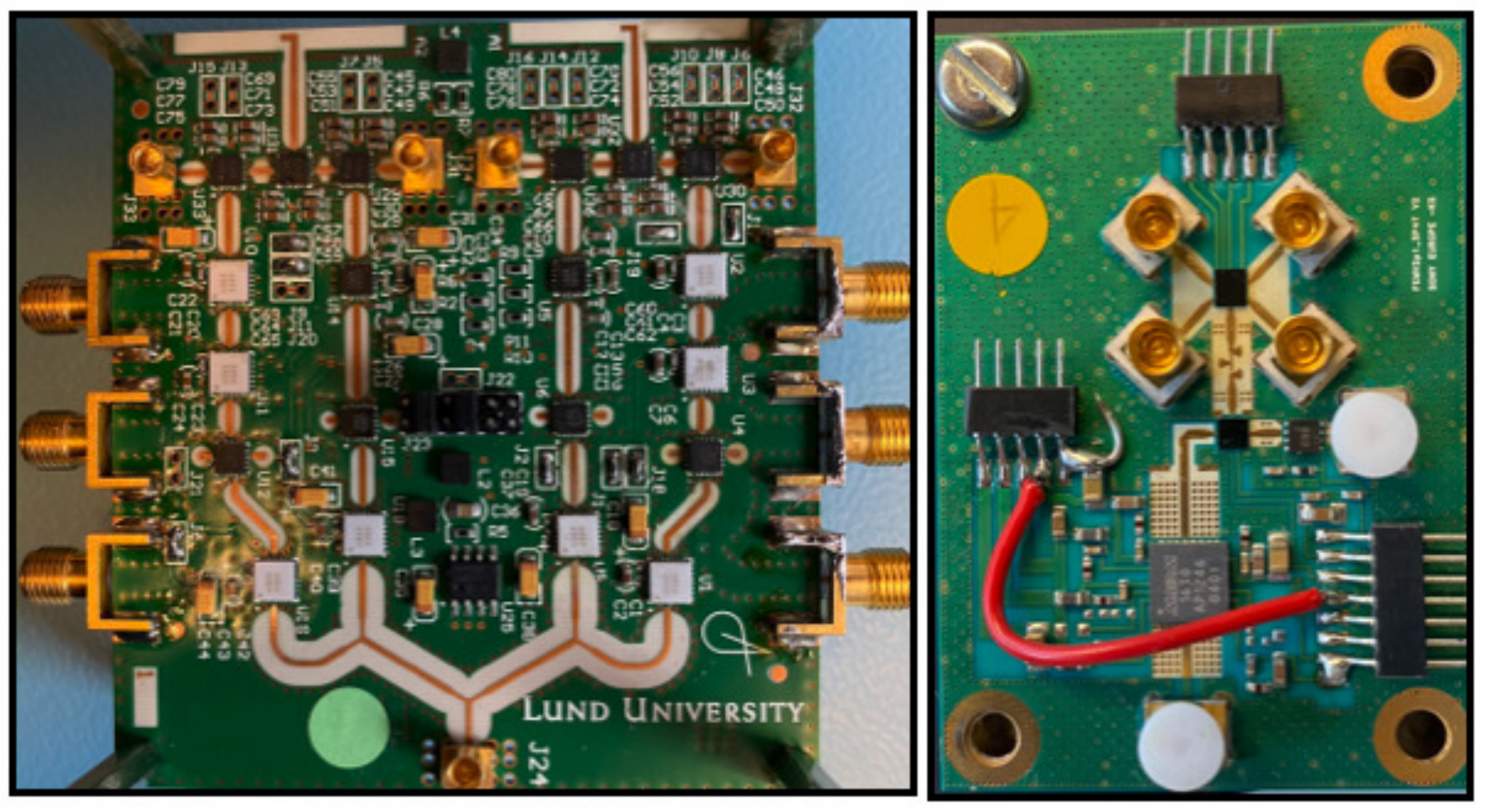}
\label{fig:photo_FRECON_FEM}
}
\caption{FRECON and FEM: (a) block diagram with one FRECON and two FEMs in an \ac{UE} (b) photographs of fabricated FRECON (left) and FEM (right).} 
\label{fig:FRECON_FEM}
\end{figure}

\section{Analog Subsystem Design and Implementation}
\label{sec:analog_sub}
\ac{RF}-interference suppression is crucial to achieve a sufficient output power of each \ac{TRx} chain. Also, beam switchability is one of key design features in the analog subsystem, to operate the proposed antenna selection algorithm in the digital subsystem. This section elaborates on the design of \ac{FRECON}, \ac{FEM}, \ac{ClkDist}, and \ac{UE} antenna arrays. Furthermore, its implementation details are discussed. 

\subsection{\ac{FRECON} and \ac{FEM}}
\label{subsec:frecon_fem}
MmWave systems are more sensitive to \ac{PA} nonlinearities, compared to conventional systems below-\SI{6}{\GHz}. For the \ac{FRECON} design, we focused on an appropriate architecture design and component selection to reduce the \ac{PA} nonlinearities. As mentioned in \scn\ref{sec:lumami28_ov}, one \ac{FRECON} has two \ac{TRx} chains, and connects with four elements of 16-element array antenna in the \ac{BS}, with two \acp{FEM} in the \ac{UE}. A combined block diagram of \ac{FRECON} and \ac{FEM} in an \ac{UE} is shown in \fig\ref{fig:BD_FRECON_FEM}. The \ac{FRECON} contains the same eight \acp{DA}~(HMC383LC4) but has different targets. Four \acp{DA} between an \ac{LO}-input port and mixers is for amplifying the \SI{25.5}{\GHz}-\ac{LO} signal, and the other four \acp{DA} for the \SI{28}{\GHz} \ac{Tx} signal. The mixers~(HMC1063LP3E) are used for up/down-conversion between \ac{IF} and \SI{28}{\GHz} bands, where an \ac{LO} power of more than $10\hspace{0.2em}\rm{dBm}$ is required to operate it. That is the reason why a \ac{DA} for amplifying the \ac{LO} signal is needed for each mixer. The conversion gain of the mixer is around \SI{-10}{\dB}. To compensate this power loss and achieve high output power, the \ac{Tx} chain is equipped with two consecutive \acp{DA}. On the other hand, each \ac{Rx} chain contains an \ac{LNA}~(HMC1040LP3CE) to avoid compression. In the front-end of the \ac{FRECON}, there are SPDT switches~(ADRF5020) for \ac{TDD} switching{\footnote{We integrated two commercial antennas for future work, together with four \ac{RF} ports. Thus, there are a total of six SPDT switches to control all the \ac{RF} inputs/outputs of \ac{FRECON}.}}. A \SI{2.45}{\GHz}-band \ac{LPF}~(SAFEA2G45) is included between \ac{Rx} mixer and \ac{IF} port.

The employment of \ac{FEM} is to support testing of long-range communications, as well as beam-switching. As depicted in \fig\ref{fig:BD_FRECON_FEM}, one \ac{FEM} contains an additional \ac{PA}~(MAAP-011246) and \ac{LNA}, which have a high power gain. The SP4T switch engages with four \ac{RF} ports, and performs switching or selecting by control signals from digital subsystem. For the \ac{TDD}\hspace{0.1em}/\hspace{0.1em}beam-switching, the isolation between paths in the switches is important. Both SPDT and SP4T switches were designed in-house using Sony's SOI process. They achieve a low insertion loss of $1.5\hspace{0.1em}{\rm dB}$ and $1.6\hspace{0.1em}{\rm dB}$, respectively, and a high isolation of $30\hspace{0.1em}{\rm dB}$ and $26\hspace{0.1em}{\rm dB}$, respectively, at \SI{28}{\GHz}. For suppressing the second harmonic in \ac{Tx} mode, and the high-frequency out-of-band spurious, including from \SI{60}{\GHz}~(WLAN\hspace{0.1em}/\hspace{0.1em}WiFi), in \ac{Rx} mode operation, an \ac{LPF} was incorporated between SPDT and SP4T. The insertion loss of this filter plays a very critical role for achieving better output power and noise-floor performance the module. A novel \ac{LPF} was designed on \ac{PCB} using distributed elements where capacitors were realized from radial stubs. This approach could realize the filter within a very compact size. It attains a very low insertion\hspace{0.1em}/\hspace{0.1em}return loss of less than $0.7\hspace{0.1em}{\rm dB}$ and $-18\hspace{0.1em}{\rm dB}$, respectively, at \SI{28}{\GHz}, and the harmonic rejection of more than $30\hspace{0.1em}{\rm dBc}$.

Based on the measurements of each module~\cite{minkeun20_11}, the \ac{Tx} and \ac{Rx} gains for the \ac{FRECON} are around \SI{9}{\dB} and \SI{7}{\dB}, respectively. For the \ac{FEM}, around \SI{14}{\dB} and \SI{12}{\dB}, respectively, is achieved. The \ac{Tx} gain in its linear region is around \SI{22}{\dB}. It delivers a 1-dB gain compression point (P1dB) of $18\hspace{0.2em}\rm{dBm}$. The maximum \ac{Rx} gain is \SI{18.8}{\dB} at \SI{27.95}{\GHz}. Also, the power consumption of the implemented \ac{FRECON} and \ac{FEM} is \SI{6.3}{\W} and \SI{7}{\W}, respectively. Photographs of the fabricated \ac{FRECON} and \ac{FEM} are shown in \fig\ref{fig:photo_FRECON_FEM}.
\begin{figure}[t!]
\centering
\includegraphics[width = 3.39in]{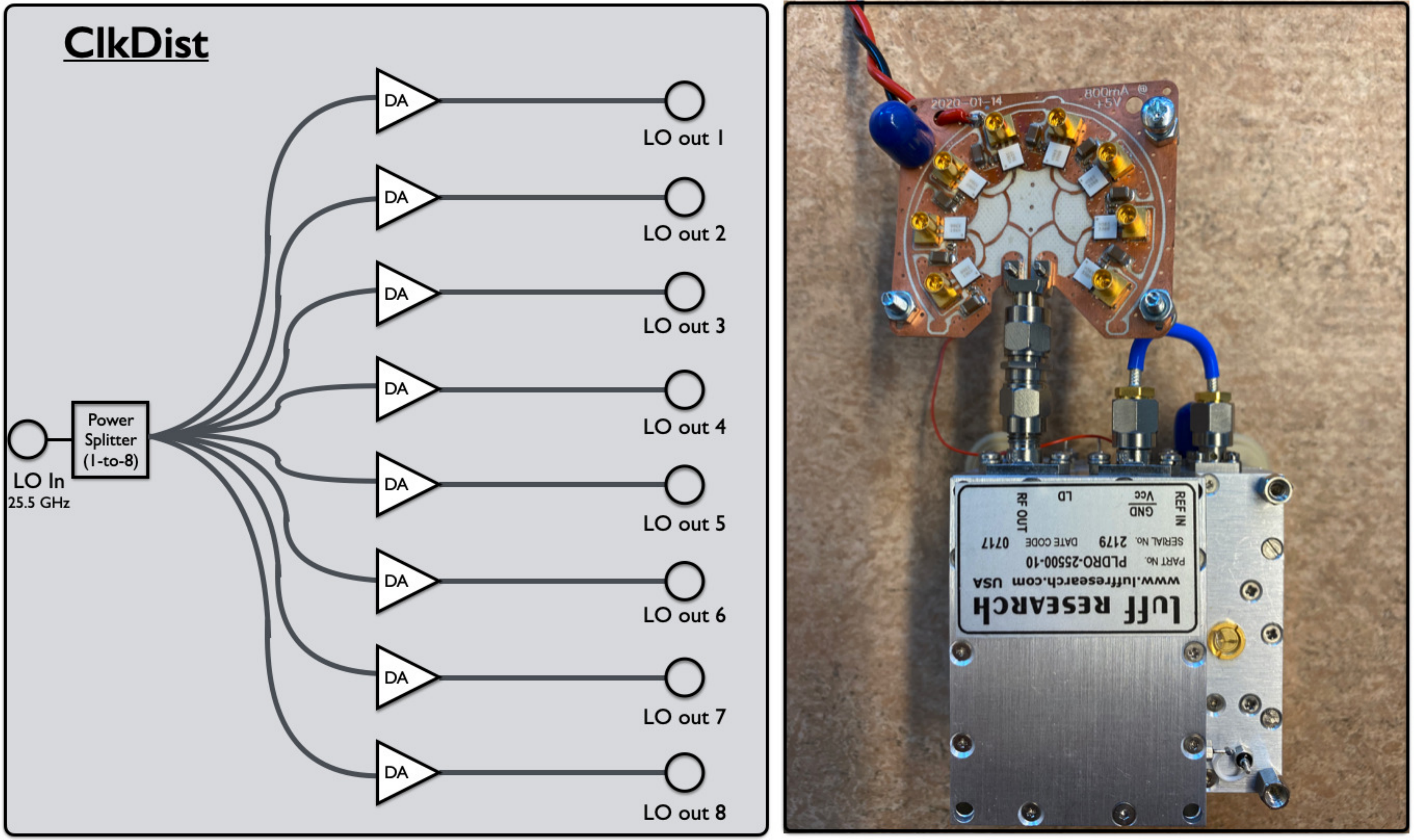}
\caption{Block diagram of ClkDist~(left) and its photograph with a common \SI{25.5}{\GHz}-LO~(right).}
\label{fig:BD_Photo_ClkDist}
\vspace{-0.25 cm}
\end{figure}

\subsection{\ac{ClkDist}}
The block diagram of the \ac{ClkDist} and its photograph are shown in \fig\ref{fig:BD_Photo_ClkDist}. The \ac{ClkDist} has 1 input and 8 output ports connecting with the \ac{LO}~(PLDRO-25500-10) and \acp{FRECON}, respectively. Since the 1-to-8 power splitter causes a power loss of more than \SI{10}{\dB}, each path in the \ac{ClkDist} is equipped with one \ac{DA}~(HMC383LC4)  to meet the input-power requirement of mixers in \ac{FRECON}. The power consumption of the fabricated \ac{ClkDist} and the \ac{LO} is \SI{4}{\W} and \SI{5}{\W}, respectively{\footnote{\acp{LO} operating at \ac{mmWave} frequencies is very sensitive to temperatures. Thus, we adopt a cooling fan to operate our \SI{25.5}{\GHz}-\ac{LO}. Its power consumption is added in the \ac{LO}'s power consumption.}}.
\begin{figure}[t!]
\centering
\subfigure[yagi antenna]
{
\includegraphics[width=3.45in]{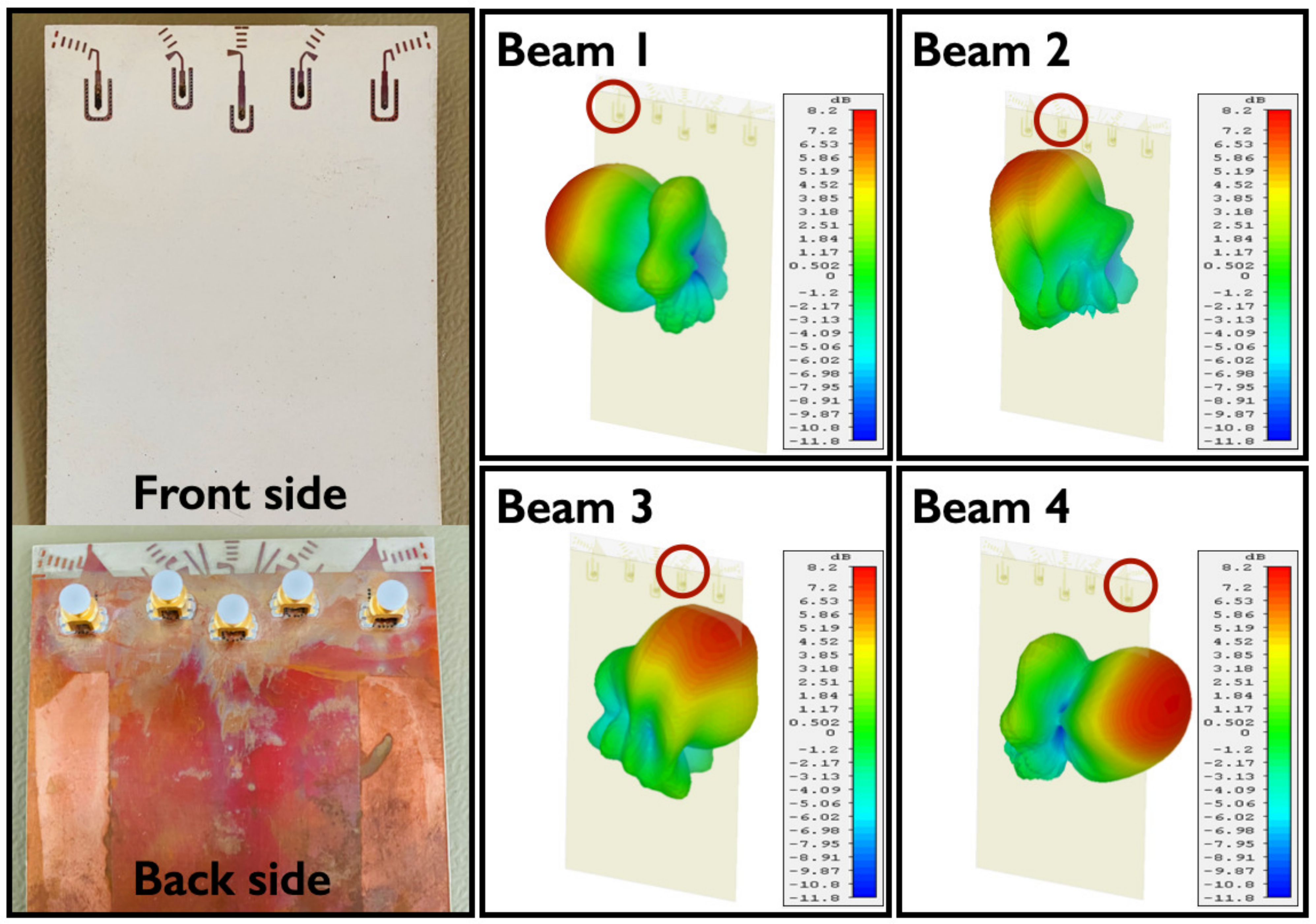}
\label{fig:yagi_ue1}
}
\vspace*{0.25 cm}
\subfigure[patch antenna]
{
\includegraphics[width=3.45in]{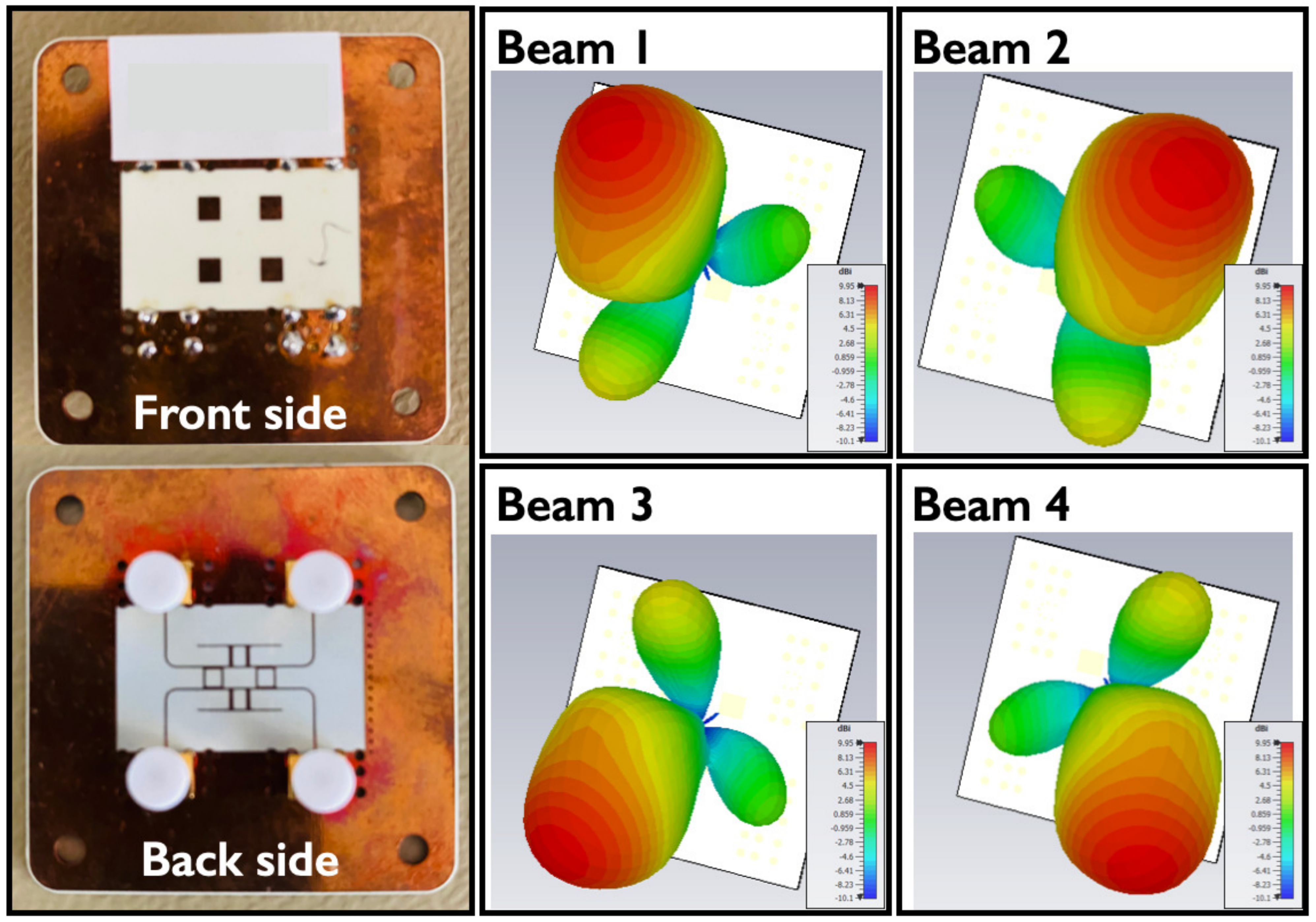}
\label{fig:patch_ue2}
}
\caption{Photograph and beam pattern of \ac{UE} antenna arrays. Four elements (out of five) of yagi antenna were used for beam selection. The peak gain of yagi antenna are $7.5\hspace{0.2em}\rm{dBi}$, $7.0\hspace{0.2em}\rm{dBi}$, $7.0\hspace{0.2em}\rm{dBi}$, and $7.5\hspace{0.2em}\rm{dBi}$, respectively, in the antenna-port order. The peak gain of patch antenna has around $10\hspace{0.2em}\rm{dBi}$ in each antenna port.} 
\label{fig:antenna_array_ue}
\end{figure}
\subsection{Antenna Array}
We designed two \ac{UE} array antennas for evaluating system performances. \fig\ref{fig:antenna_array_ue} shows photographs and beam patterns of two array antennas. \fig\ref{fig:yagi_ue1} is a wideband~(\SI{26}{\GHz} to \SI{40}{\GHz}) Yagi antenna~\cite{dipaola19} with multiple beam points to different directions, which forms a wide angle coverage. It is designed on a two-layer \ac{PCB} using RO3003 substrate. Four elements were used for beam selection, where each antenna gain is, respectively, $7.5\hspace{0.2em}\rm{dBi}$, $7.0\hspace{0.2em}\rm{dBi}$, $7.0\hspace{0.2em}\rm{dBi}$, and $7.5\hspace{0.2em}\rm{dBi}$. \fig\ref{fig:patch_ue2} show a $2 \times 2$ patch antennas with a butler matrix, capable of forming four directional beams. It is designed on a three-layer \ac{PCB} using two stacked RO4350B substrates. The antenna-element spacing in the subarray is half a wavelength~($\lambda/2$), \ie, \SI{5.5}{\mm}. The peak antenna gain and its bandwidth are around $10\hspace{0.2em}\rm{dBi}$ and \SI{1.5}{\GHz}. Both antennas have linear polarization.

\section{Measurement Results}
\label{sec:measure}
This section uses measurement results from static and mobility environnments to study the system performance of LuMaMi28. The main measurement campaigns are as follows. First, a path loss measurement with yagi and patch antennas was carried out, where the measured path losses are compared to theoretical ones. Second, a real-channel data between LuMaMi \ac{BS} and \acp{UE} was captured to explore potential gains, achieved by the proposed antenna-selection algorithm. Lastly, real-time performance of LuMaMi28 in a variety of mobility environments was measured. According different digital-beamforming algorithms and different \ac{UE} antennas, the LuMaMi28 performance results will be presented. In the next subsection, we will describe our measurement scenarios followed by the performance details. 
\begin{figure*}[t!]
\centering
\subfigure[ ]
{
\includegraphics[width=3.99in]{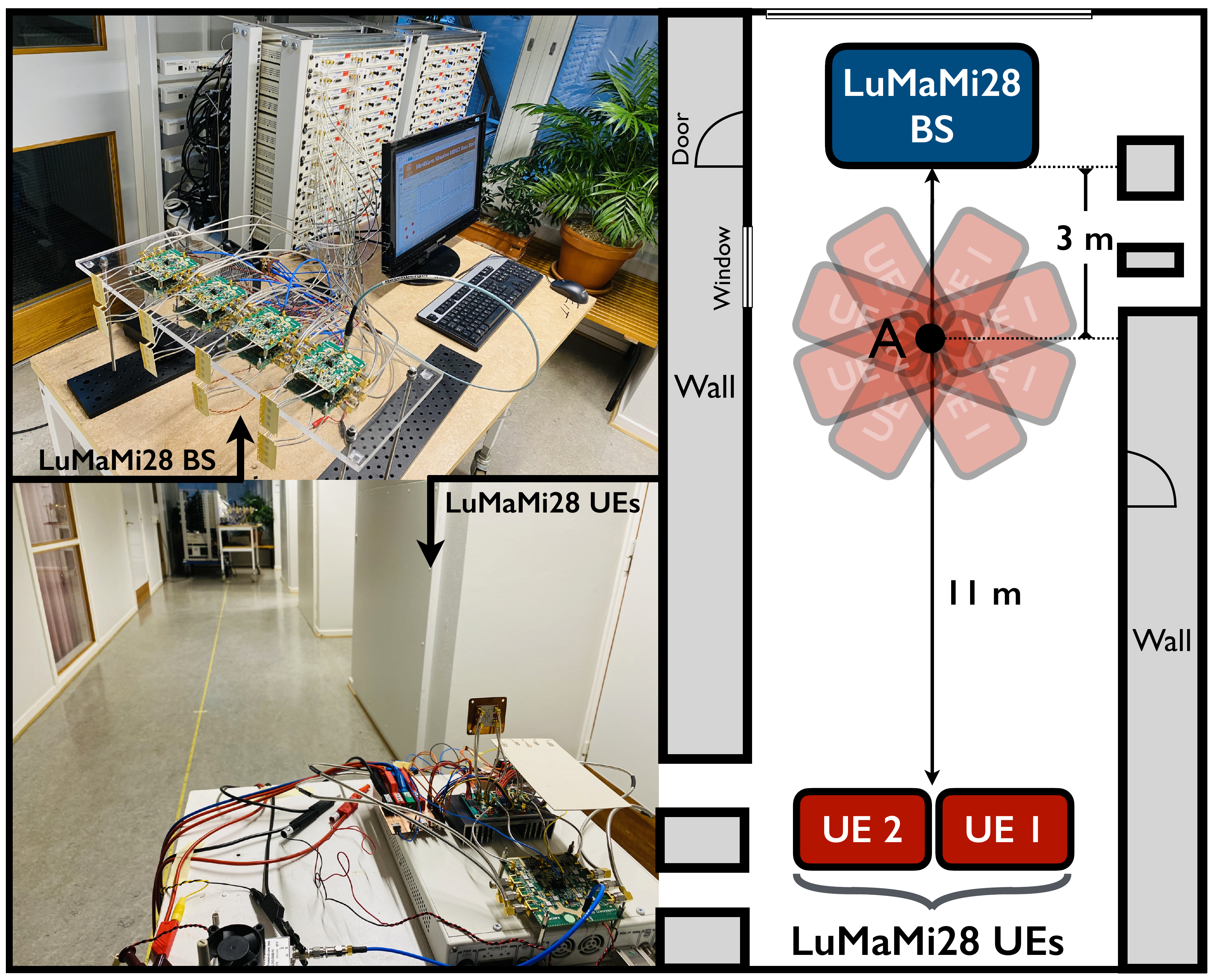}
\label{fig:static_env}
}
\subfigure[ ]
{
\includegraphics[width=2.89in]{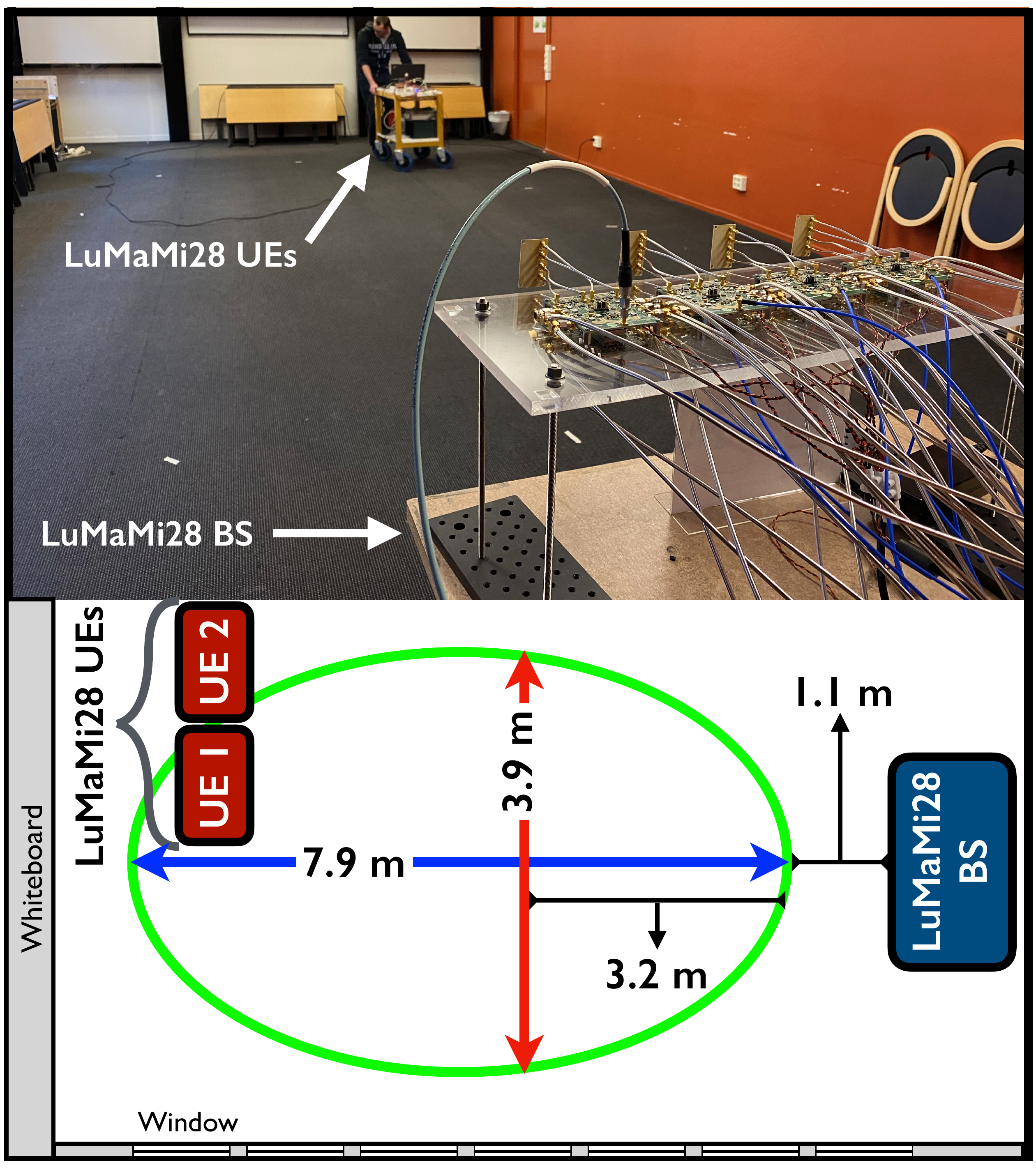}
\label{fig:mobility_env}
}
\caption{Photos and indoor maps of measurement campaign in static\hspace{0.2em}/\hspace{0.2em}mobility environments. There is no obstruction between the LuMaMi28 BS and UEs in all the environments. Both static and mobility measurement places are surrounded by concrete wall and clear windows. (a) setup for the static environment at a long corridor in E-huset buildng. Spot A is the location for \ac{UE} rotation test. (b) setup for the mobility environments at a large lecture hall in the same building. There are three routes for its measurement campaigns; circled~(green), vertical~(blue), and horizontal routes~(red).} 
\label{fig:mea_scenario}
\vspace*{-0.25 cm}
\end{figure*}

\subsection{Measurement Scenarios}
\subsubsection{LuMaMi28 Setup}
A LuMaMi28 \ac{BS} and two \acp{UE} were used, where the \ac{BS} and each \ac{UE} have 16-\ac{TRx} chains and a 4-antenna\hspace{0.15em}/\hspace{0.15em}single-\ac{TRx} unit, respectively, as shown in \fig\ref{fig:Arch_testbed}. All the measurement campaigns were performed in E-huset building at Lund University. \fig\ref{fig:mea_scenario} shows photos and indoor maps of our measurement campaign. We selected a long corridor and a large lecture hall for the static and mobility experiments, respectively. In the corridor, the distance between the \ac{BS} and \acp{UE} ranged from \SI{3}{\m} to \SI{11}{\m}, as depicted in \fig \ref{fig:static_env}. Here, a path loss measurement, an \ac{UL} channel data recording, and an \ac{UE} rotation test were performed. We will provide more details on measurement results in the long corridor in \scn \ref{subsec:path_loss} -- \scn \ref{subsec:rotation_test}. In the lecture hall, we performed mobility tests with three different routes, as shown in \fig \ref{fig:mobility_env}. The BS looked at the whiteboard at a fixed spot. Based on the \ac{BS} location, the green line is for the circled route, the blue for the vertical route, and the red for the horizontal route. Two \acp{UE} were co-located on a moving cart. This cart moved with a pedestrian speed of around $3\hspace{0.2em}\rm{km/h}$ in each route. The transmit power of both \acp{UE} is fixed over all the measurement scenarios. \fig \ref{fig:ul_power_spec} shows a captured \ac{UL} \ac{Tx} power spectrum at a \ac{FEM} \ac{Tx} port. We will present measurement details in the lecture room in \scn\ref{subsec:mobility_test}.

\subsubsection{Metrics}
our evaluation uses \ac{Rx} power for path loss measurement, channel gain for \ac{UL} channel capturing, and \ac{UL} throughput and throughput gain for mobility test. The individual throughput of each \ac{UE} were recorded in real-time. The throughput gain is computed as the ratio of the measured \ac{UL} throughputs with and without antenna selection. 
\begin{figure}[t!]
\centering
\includegraphics[width = 3.39in]{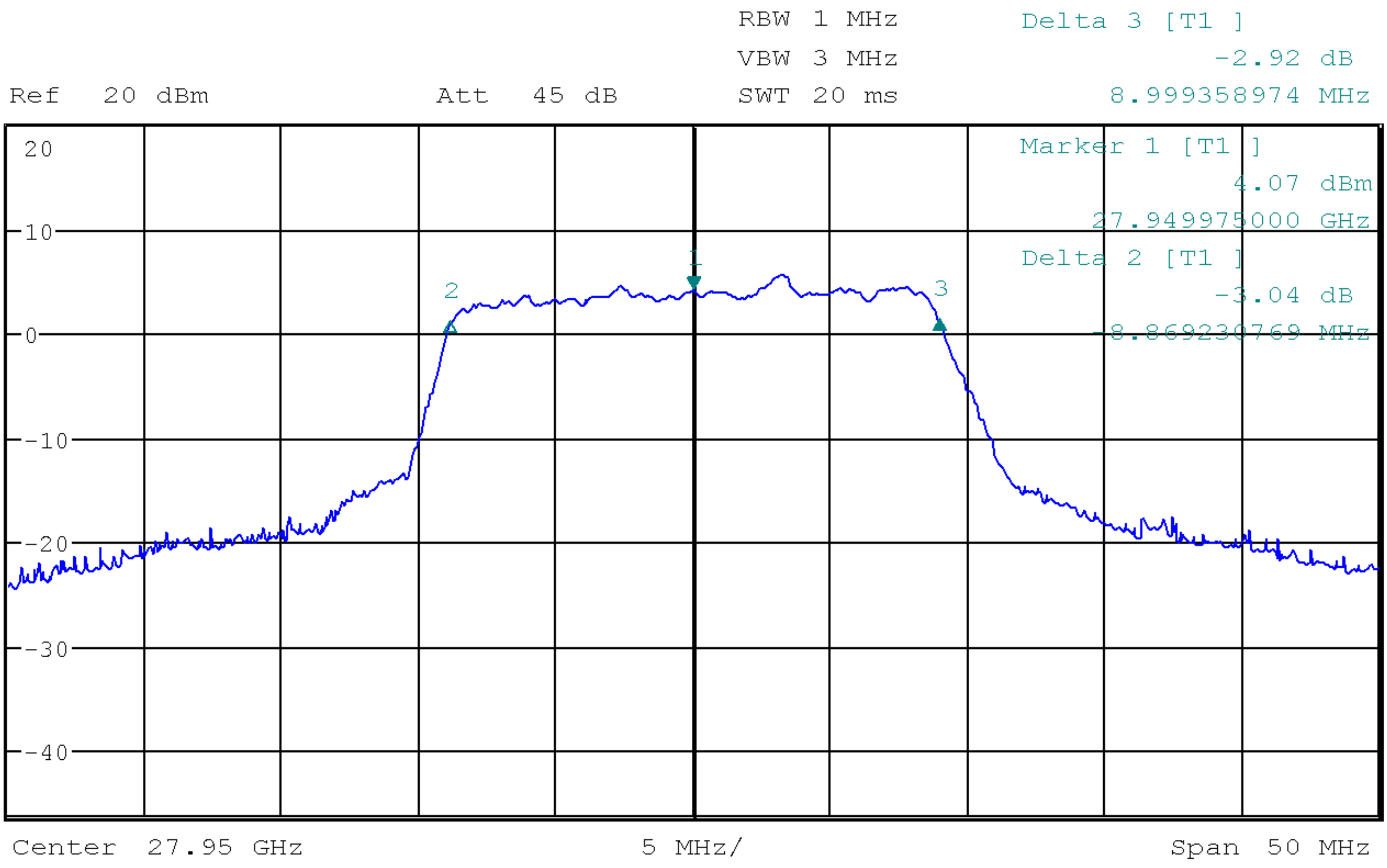}
\caption{An \ac{UL} \ac{Tx} power spectrum captured at a \ac{FEM} \ac{Tx} port. Max hold mode was used for peak detection. In this captured spectrum, the peak power is $4.07\hspace{0.2em}\rm{dBm}$ at the center frequency of \SI{27.95}{\GHz}.}
\label{fig:ul_power_spec}
\vspace{-0.25 cm}
\end{figure}

\subsubsection{Antenna Type} 
the \ac{UE}1 was equipped with the yagi antenna~(\fig \ref{fig:yagi_ue1}), and the \ac{UE}2 equipped with the patch antenna~({\fig \ref{fig:patch_ue2}). This configuration is identical over all the measurement campaigns. During the \ac{UL} throughput measurements, two co-located \acp{UE} moved together to provide the same mobility environment to both \acp{UE} as possible.

\subsubsection{Equalization}
the LuMaMi28 BS performs a fully-digital beamforming. In the measurement campaign, the \ac{MRC} and \ac{ZF} equalizers in (\ref{eq:linear_eq}) were used to decode \ac{UL} signals from two \acp{UE}.

\subsection{Path Loss Measurement}
\label{subsec:path_loss}
Prior to the performance measurement, we conducted the \ac{Rx} power measurement of the LuMaMi28 \ac{BS} for link-budget calculation. A signal generator~(E8257D) was used, as an \ac{IF} signal generator, to guarantee an accurate power for transmission. The output of E8257D was connected with one \ac{FRECON}, one \ac{FEM}, and one antenna in order, \ie, only the \ac{SDR} part was replaced with E8257D from an \ac{UE} setup. Both yagi and patch \ac{UE} antennas were employed for the link-budget calculation. The \ac{IF} \ac{Rx} powers of 16 \acp{FRECON} in the LuMaMi28 \ac{BS} were measured by using a spectrum analyzer~(FSU50). We selected the following distances~($\rm{m}$), $\mathcal{D}_{\mathsf{p}} = \{ 3, 5, 7, 9, 11\}$. At each \ac{Tx} and \ac{Rx} locations, the \ac{Rx} power measurements were conducted with boresight to the center of the LuMaMi28 \ac{BS} array\footnote{The array antennas of \ac{BS} and \acp{UE} were aligned with co-polarization. The same antenna setup was applied for all experiments.}.  
\begin{figure}[t!]
\centering
\includegraphics[width = 3.39in]{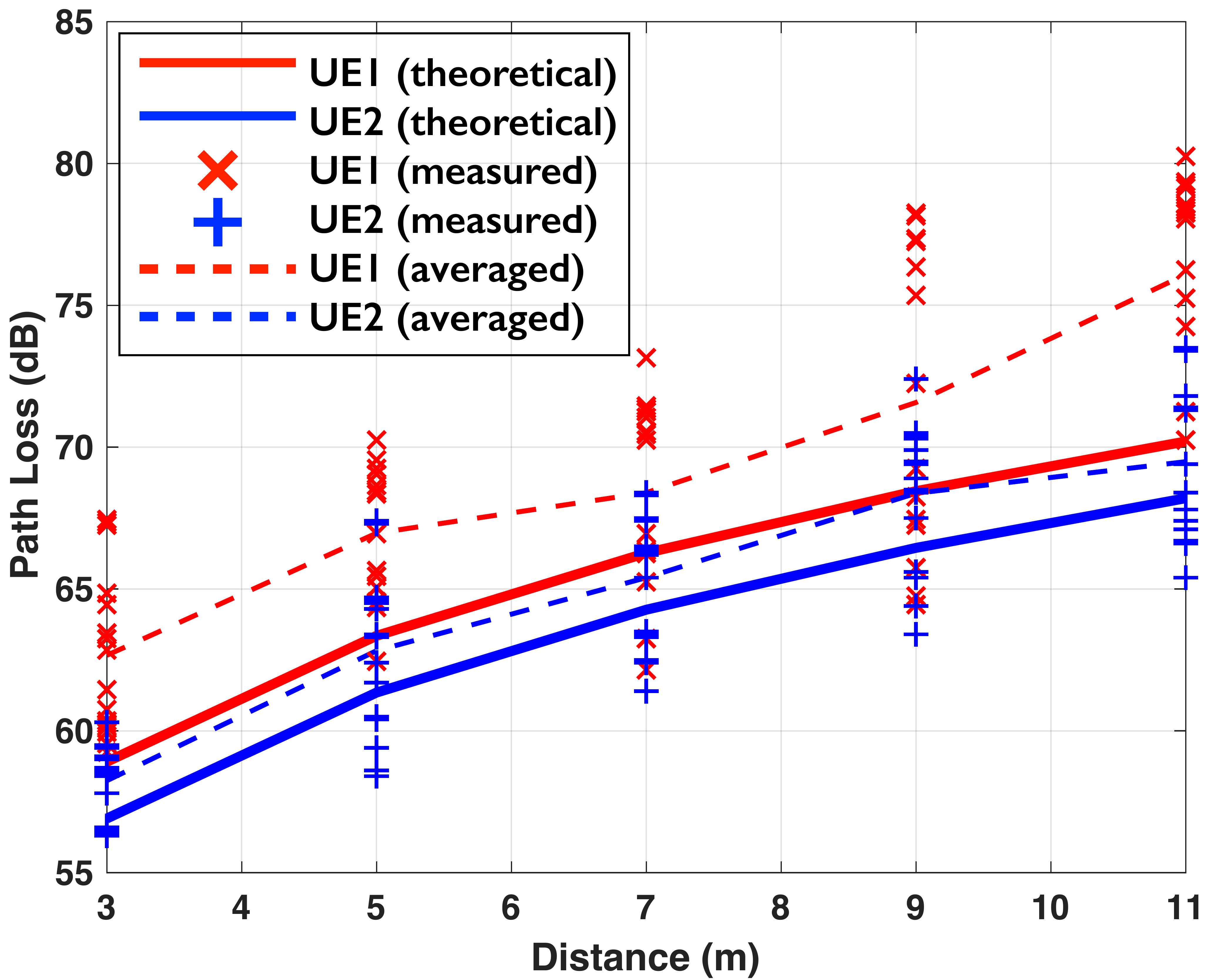}
\caption{Measured path loss values of \ac{UE}1~(yagi)\hspace{0.2em}/\hspace{0.2em}\ac{UE}2~(patch) for \SI{27.95}{\GHz} indoor channels. At each distance, the \acp{FSPL} measured from 16 \ac{FRECON} are represented. The dotted red~(\ac{UE}1)\hspace{0.2em}/\hspace{0.2em}blue~(\ac{UE}2) lines are the averaged values of the measured path loss. }
\label{fig:PL_measurement}
\vspace{-0.25 cm}
\end{figure}

Based on the measured \ac{Rx} power values, we calculated a measured \ac{FSPL} for the distance $d$ between \ac{Tx} and \ac{Rx} antennas, and compare with its theoretical number. \ac{EIRP} is the  hypothetical power radiated by a isotropic \ac{Tx} antenna in the strongest direction and defined as
\begin{align}
\label{eq:eirp}
\mathsf{EIRP}({\rm{dBm}})
= 
P_{\mathsf{tx}}
+
G_{\mathsf{tx}}^{\mathsf{c}}
-
L_{\mathsf{tx}}^{\mathsf{c}}
+
G_{\mathsf{tx}}^{\mathsf{a}}	
\end{align}
where $P_{\mathsf{tx}}$ is the \ac{Tx} power~(\ac{IF} input), $G_{\mathsf{tx}}^{\mathsf{c}}$ the effective \ac{Tx} gain of \ac{FRECON} and \ac{FEM}, $L_{\mathsf{tx}}^{\mathsf{c}}$ the cable loss in the \ac{Tx} side, and $G_{\mathsf{tx}}^{\mathsf{a}}$ the \ac{Tx} antenna gain.
Using the \ac{EIRP}, the measured \ac{FSPL} is
\begin{align}
\label{eq:mPL}
\mathsf{PL}_{\mathsf{m}}({\rm{dB}})
= 
\mathsf{EIRP}
-
(P_{\mathsf{rx}}
-
G_{\mathsf{rx}}^{\mathsf{c}}
+
L_{\mathsf{rx}}^{\mathsf{c}}
-
G_{\mathsf{rx}}^{\mathsf{a}}	)
\end{align}
where $P_{\mathsf{rx}}$ is the \ac{Rx} power~(\ac{IF} output), $G_{\mathsf{rx}}^{\mathsf{c}}$ the effective \ac{Rx} gain of \ac{FRECON}, $L_{\mathsf{rx}}^{\mathsf{c}}$ the cable loss in the \ac{Rx} side, and $G_{\mathsf{rx}}^{\mathsf{a}}$ the \ac{Rx} antenna gain. The theoretical \ac{FSPL} is
\begin{align}
\label{eq:tPL}
\mathsf{PL}_{\mathsf{th}}({\rm{dB}})
= 
20{\rm{log}}_{10}
\Bigg(
\frac{4\hspace{0.06em}\pi\hspace{0.06em}d\hspace{0.06em}f}
{c}
\Bigg)
-
G_{\mathsf{tx}}^{\mathsf{a}}
-
G_{\mathsf{rx}}^{\mathsf{a}}
\end{align}
where $f$ is the carrier frequency, and $c$ is the speed of light. 
\fig\ref{fig:PL_measurement} shows the measured \ac{FSPL} values of \ac{UE}1~(yagi)\hspace{0.2em}/\hspace{0.2em}\ac{UE}2~(patch) for \SI{27.95}{\GHz} indoor channels, from (\ref{eq:eirp})-(\ref{eq:tPL}). Also, the average curves of the measured \acp{FSPL} for each \ac{UE} are shown. It is shown that the both averaged \acp{FSPL} are generally quite close to the the theoretical ones. The \ac{UE}1's gap between averaged and theoretical \ac{FSPL} is relatively larger than that of the \ac{UE}2. It is conjectured that the yagi antenna beam is relatively more sensitive to placement or \ac{AoD} although a boresight measurement was applied for measuring the \ac{Rx} power at the \ac{BS}.

\begin{figure}[t!]
\centering
\subfigure[ ]
{
\includegraphics[width=3.39in]{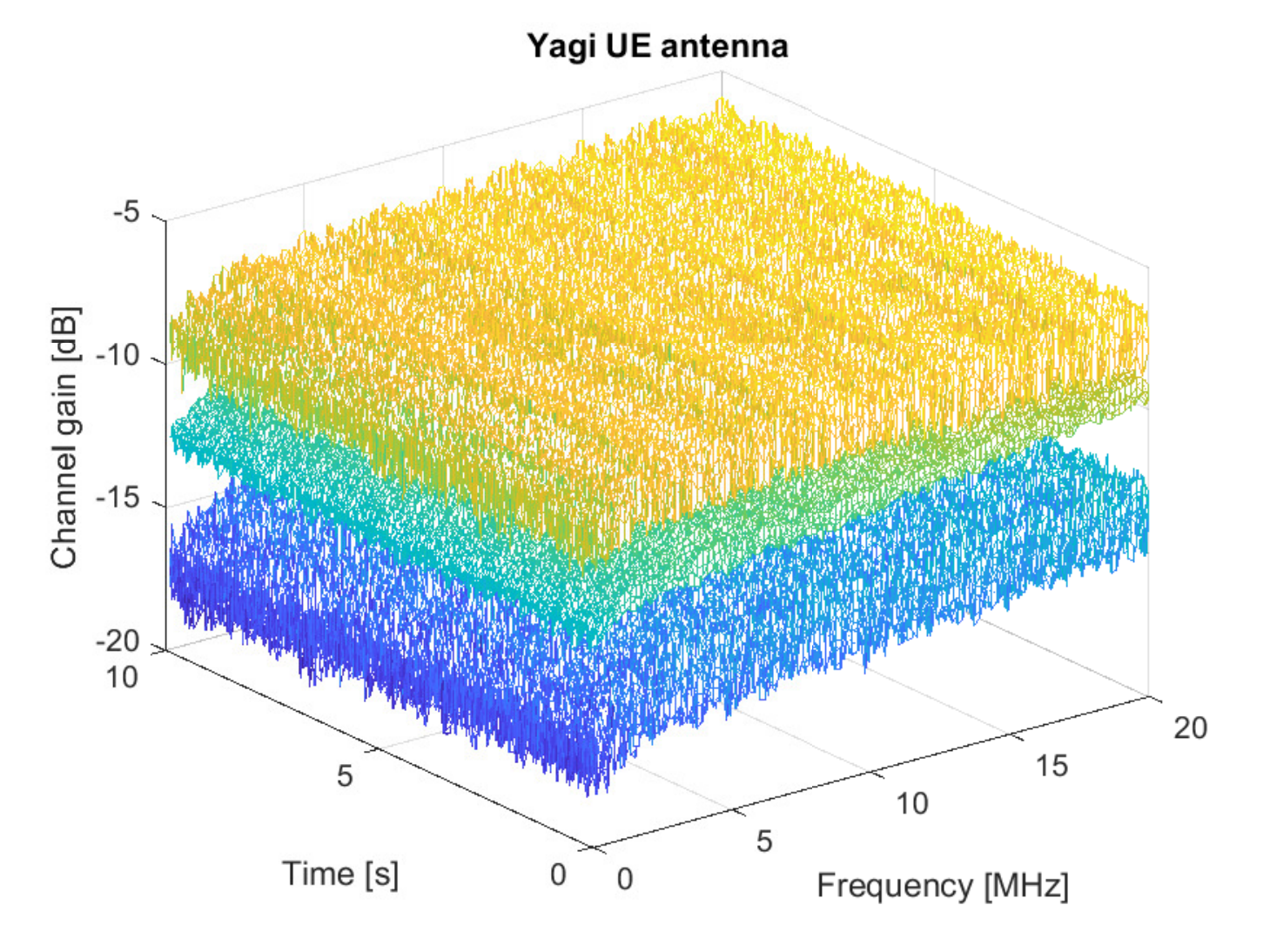}
\label{fig:yagi-3d}
}
\subfigure[ ]
{
\includegraphics[width=3.39in]{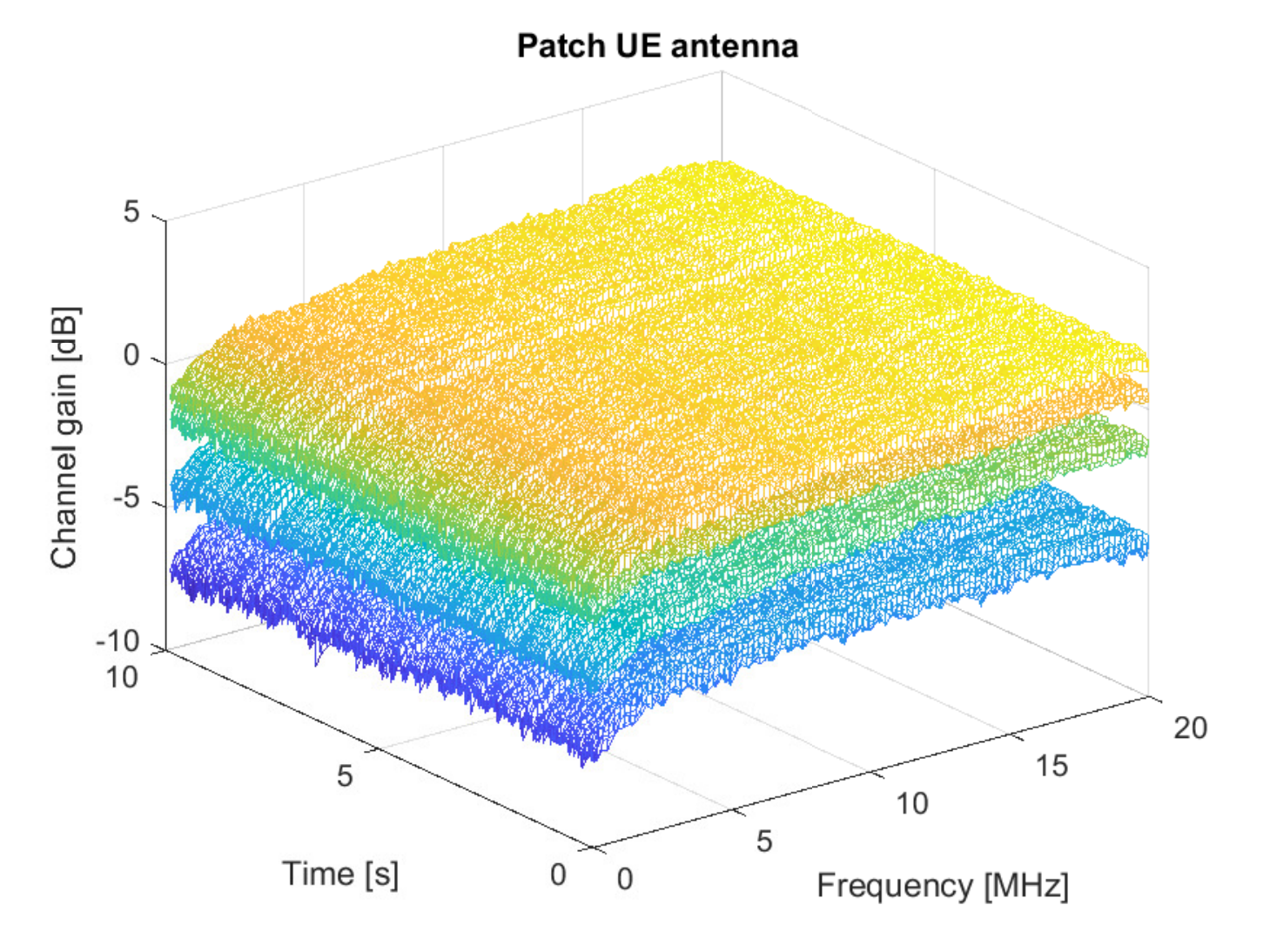}
\label{fig:patch-3d}
}
\caption{Channel gain for all four antennas over time and frequency, summed for all BS antennas (a) yagi antenna (b) patch antenna} 
\label{fig:3d-all-beams}
\end{figure}
\begin{figure*}[t!]
\centering
\subfigure[ ]
{
\includegraphics[width=3.39in]{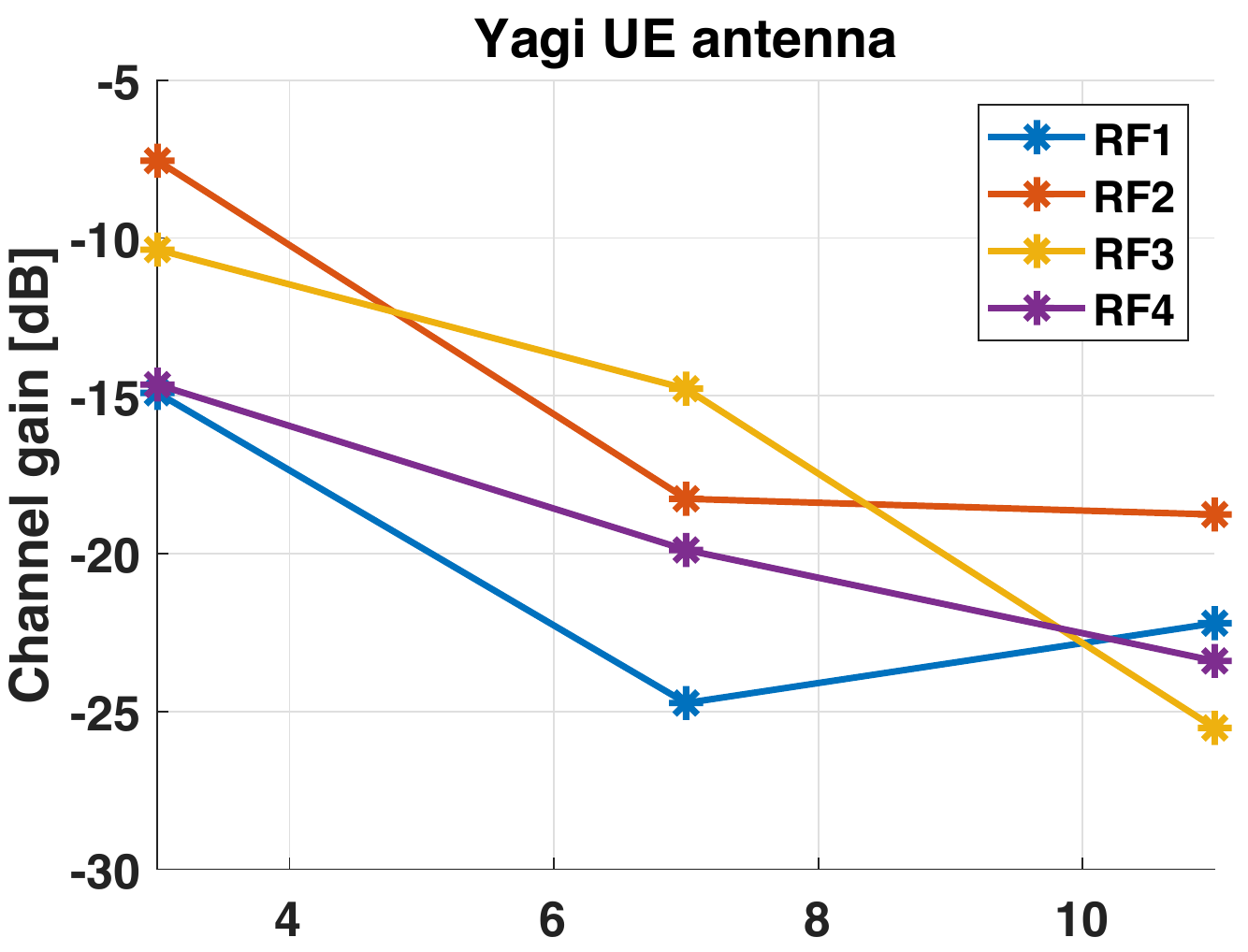}
\label{fig:yagi-distance}
}
\subfigure[ ]
{
\includegraphics[width=3.39in]{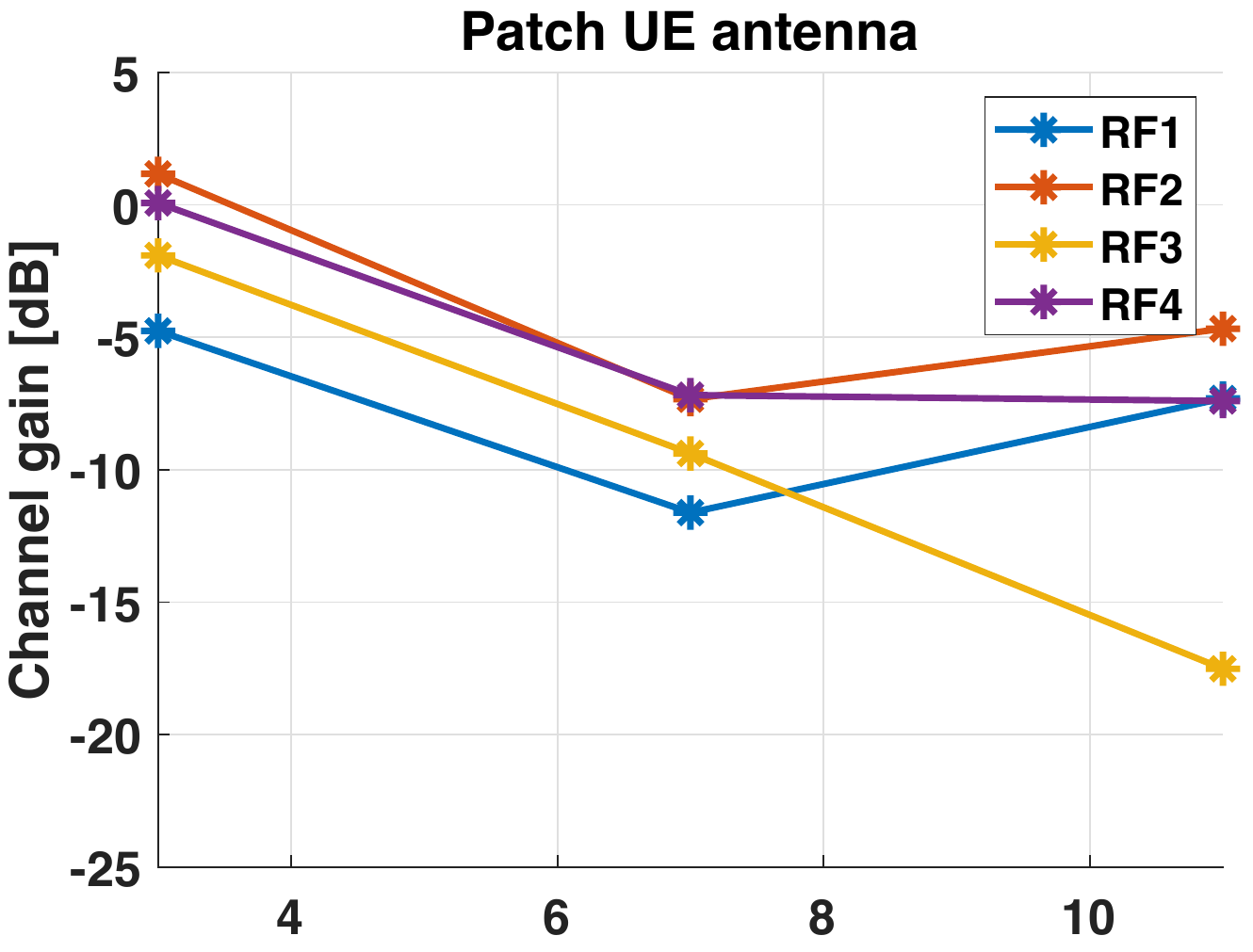}
\label{fig:patch-distance}
}
\caption{Channel gain for all four antennas over time and frequency, summed for all BS antennas (a) yagi antenna (b) patch antenna} 
\label{fig:distance-all-beams}
\vspace*{-0.25 cm}
\end{figure*}

\subsection{\ac{UL} channel capturing}
\label{subsec:ul_ch_cap}
This subsection studies potential gains achieved by the proposed antenna-selection algorithm. To better assess the potential gain, real-channel capture experiments were conducted where the captured data is the \ac{UL} channels between LuMaMi28 \ac{BS} and \acp{UE}. This was done by a flexible implementation at the \ac{BS} side, which allows for capturing the signal from the \ac{UL} pilots in real-time and then saving the channel to be used for further analysis in the post-processing. Static measurements were done with two \acp{UE} at a distance of 3, 7, and \SI{11}{\m} from the \ac{BS}. \ac{UE}1 and \ac{UE}2 have the yagi and patch antenna array, respectively. The \ac{UL} channels from the two \acp{UE} were collected every \SI{10}{\ms} for a total duration of \SI{10}{\s}. For each position, the channels were collected four times, one time for each of the antennas ports in each \ac{UE} antenna array, such that the potential gain, achieved by switching among the four antenna ports, later could be assessed by comparing the channel gain, \ie, the squared amplitudes of the channel coefficients, for each of the antenna ports.


\begin{figure*}[t!]
\centering
\subfigure[ ]
{
\includegraphics[width=3.39in]{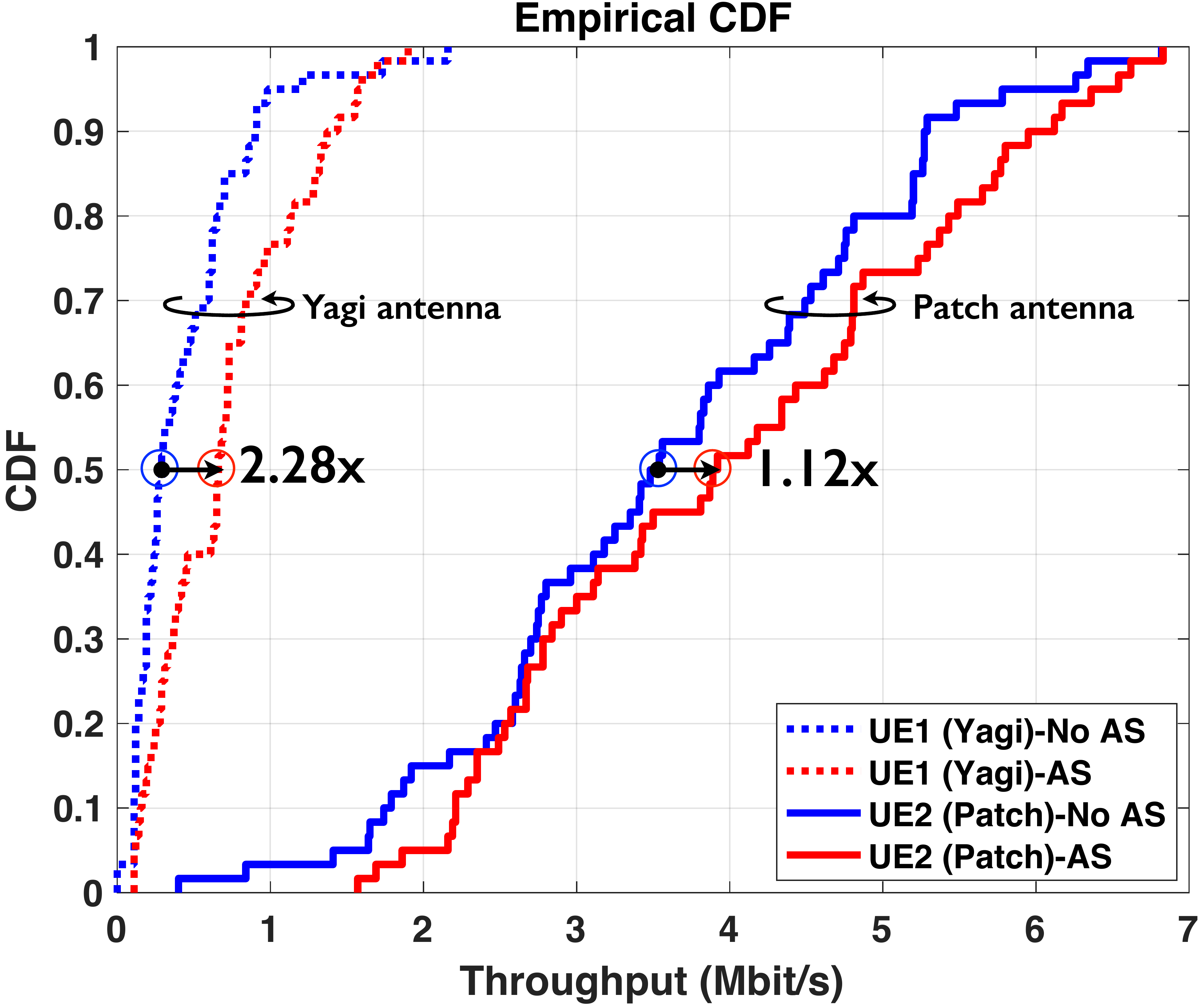}
\label{fig:15_mrc}
}
\subfigure[ ]
{
\includegraphics[width=3.39in]{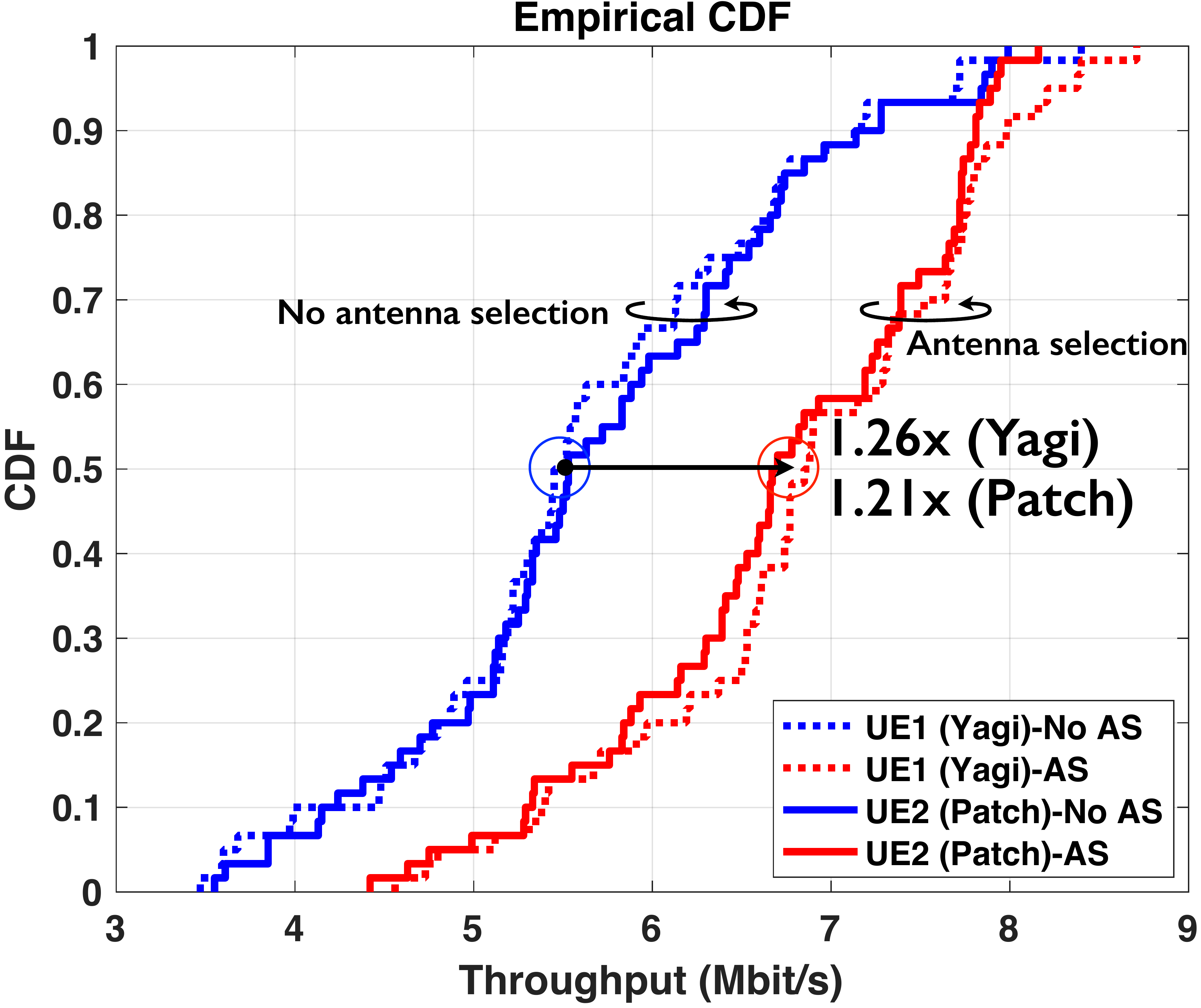}
\label{fig:15_zf}
}
\caption{Throughput \acp{CDF} of \acp{UE} rotating between $-90^\circ$ and $90^\circ$ at a fixed spot. The dotted\hspace{0.2em}/\hspace{0.2em}solid lines indicate the \ac{CDF} of the \ac{UE}1 with yagi antenna and the \ac{UE}2 with patch antenna, respectively. The red\hspace{0.2em}/\hspace{0.2em}blue curves are for antenna selection\hspace{0.2em}/\hspace{0.2em}no antenna selection, respectively. (a) MRC (b) ZF.} 
\label{fig:15_rotation}
\vspace*{-0.25 cm}
\end{figure*}

The channel gain over time and frequency, as summed over the \ac{BS} \ac{RF} chains, is shown in \fig\ref{fig:yagi-3d} for the yagi antenna array and in \fig\ref{fig:patch-3d} for the patch antenna array. These captures are at the \SI{3}{\m} distance from the \ac{BS} and each of the levels corresponds to one antenna port in the array. What can be seen in the figures are that in time the channel response is flat, as expected for a static measurement. In frequency it is close to flat, although slightly decaying for the lower part of the bandwidth. In \fig\ref{fig:yagi-3d} the channel response varies more between the samples and the total channel gain is lower than what is seen in \fig\ref{fig:patch-3d}, where the response does not vary as much between the samples. These observations are in line with the fact that the yagi antenna has a wider bandwidth and lower gain, while the patch antenna has a more narrow bandwidth and higher gain. In \fig\ref{fig:yagi-3d}, it can be observed that one antenna port clearly has a higher channel gain than the others. There is also a clear second strongest antenna while the two last ones are overlapping and can hence not be visually separated from this figure. In \fig\ref{fig:patch-3d}, the channel gain of the four antenna ports are easier to distinguish between. Overall, it can be concluded that there definitely are differences in terms of channel gain depending on the chosen antenna port and hence, there is indeed a gain to be achieved by switching between them.

To further investigate the gain that can be achieved by switching between the \ac{UE} antenna ports, the average channel gain at different distances for all the antenna ports can be seen in \fig\ref{fig:yagi-distance} for the yagi antenna array and in \fig\ref{fig:patch-distance} for the patch antenna array. The channel gain here is averaged over time and frequency and summed for all \ac{BS} \ac{RF} chains. During the measurements, depending on the distance and \ac{UE} antenna port measured, parts of the captures were lost. It is probably due to lost synchronization, influenced by weak signal at a moment. These parts were removed from further analysis. It is seen in \fig\ref{fig:distance-all-beams} that again the higher gain of the patch antenna translates into higher channel gains, compared to the yagi antenna. Looking at each measured position, one can observe that in most cases the channel to one antenna port is significantly stronger than the other three. However, it is not necessarily the same at another position. Translating this to numbers in terms of achievable gain by using \ac{UE}-antenna steering, the gains for the yagi antenna at 3, 7, and \SI{11}{\m} are between $2.8 - 7.4$, $3.5 - 10.0$, and $3.5 - 6.8\hspace{0.2em}\rm{dB}$, respectively. For the patch antenna, the corresponding gains are between $1.1-5.9$, $0 - 4.4$, and $2.6 - 12.8\hspace{0.2em}\rm{dB}$, respectively. It infers that selecting an \ac{RF} port with the highest channel magnitude can provide a significant benefit for indoor \ac{mmWave} environment. The following subsections show the actual benefit of antenna-selection , operated in LuMaMi28, in various indoor scenarios.

\begin{figure*}[t!]
\centering
\subfigure[ ]
{
\includegraphics[width=3.39in]{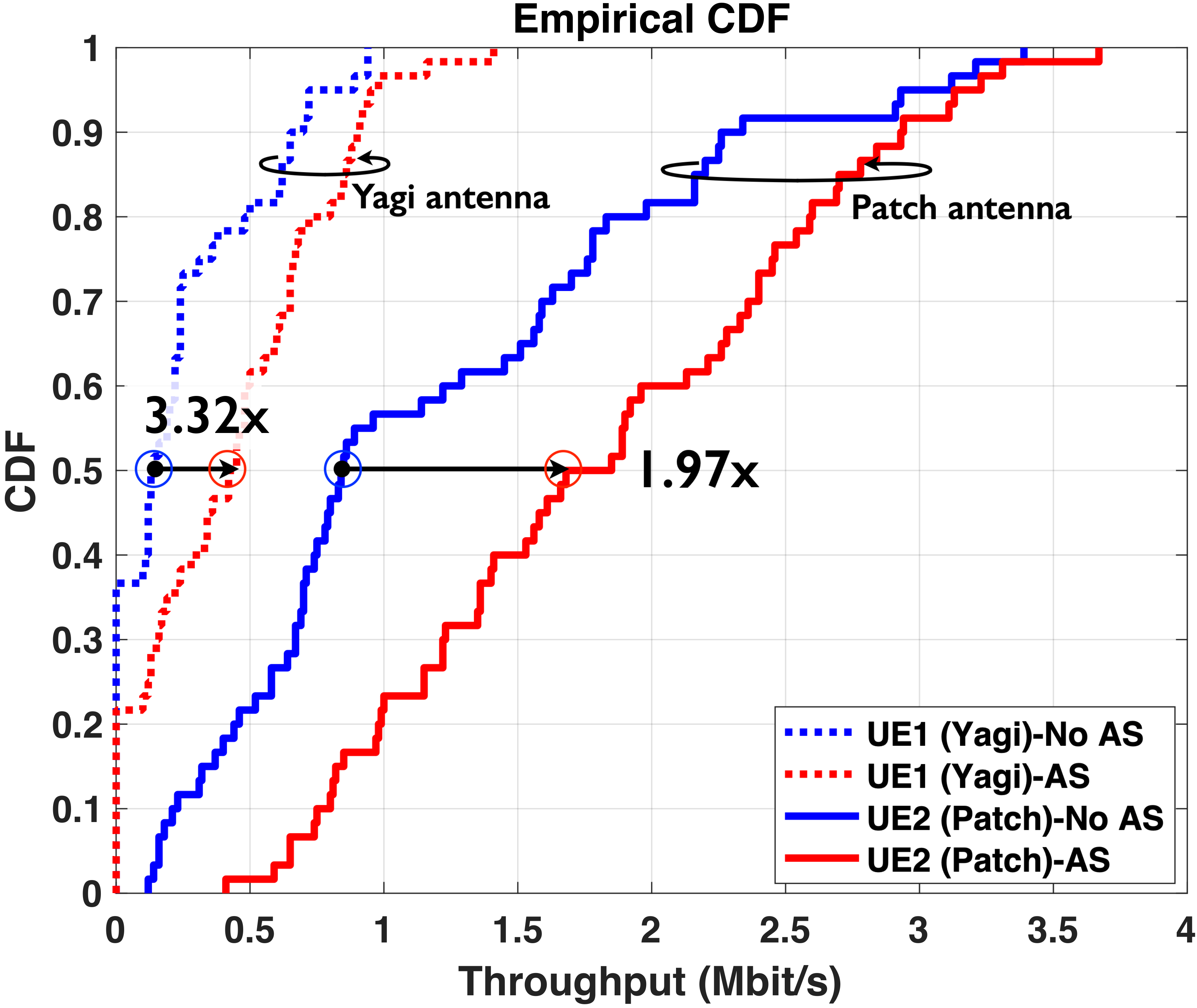}
\label{fig:16_mrc}
}
\subfigure[ ]
{
\includegraphics[width=3.39in]{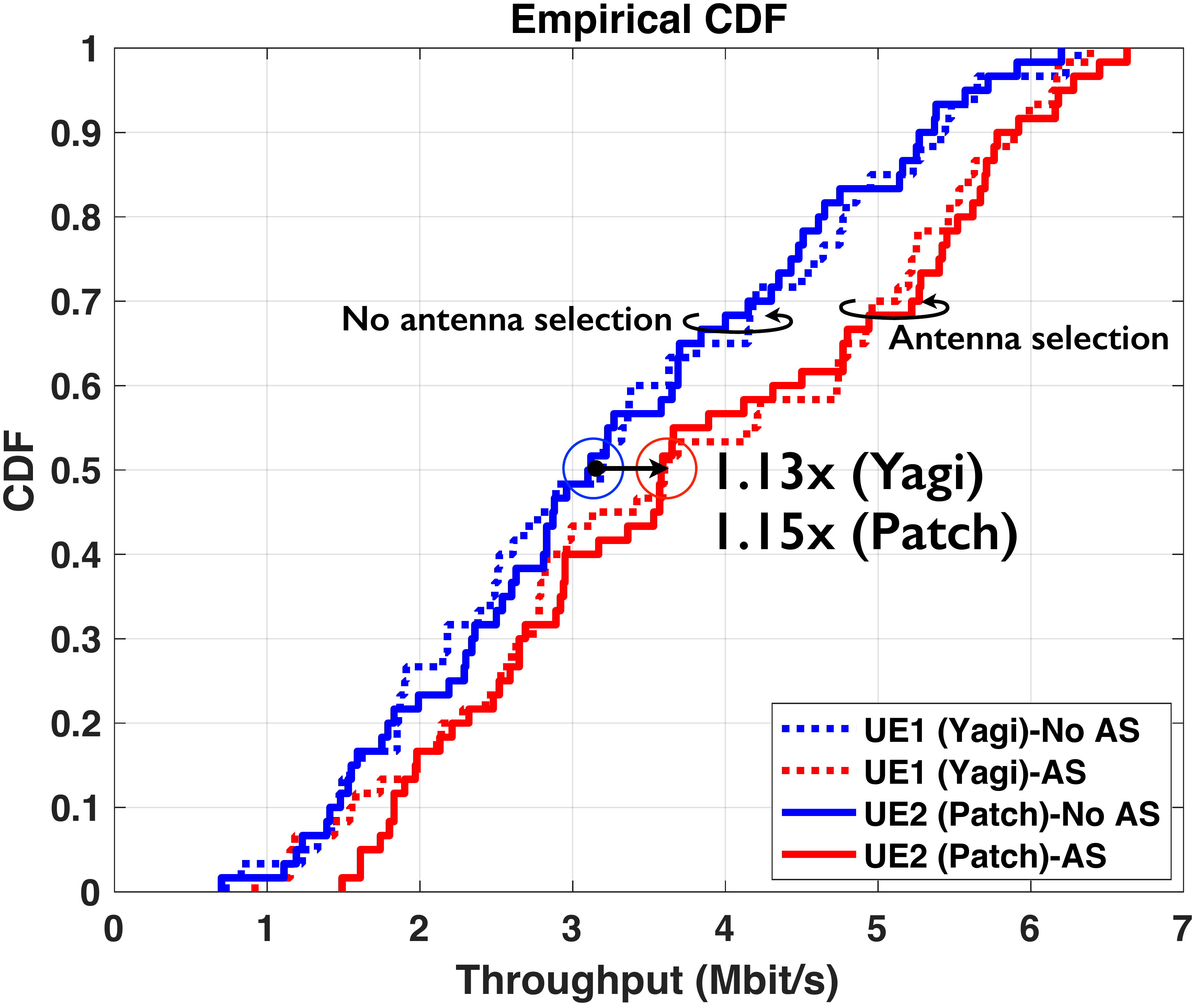}
\label{fig:16_zf}
}
\caption{Throughput \acp{CDF} of \acp{UE} moving in the horizontal route~(around $3\hspace{0.2em}\rm{km/h}$). The dotted\hspace{0.2em}/\hspace{0.2em}solid lines indicate the \ac{CDF} of the \ac{UE}1 with yagi antenna and the \ac{UE}2 with patch antenna, respectively. The red\hspace{0.2em}/\hspace{0.2em}blue curves are for antenna selection\hspace{0.2em}/\hspace{0.2em}no antenna selection, respectively. (a) MRC (b) ZF.}
\label{fig:16_horizontal}
\end{figure*}

\subsection{\ac{UE} Rotation Test}
\label{subsec:rotation_test}
We would like to verify an actual throughput gain of LuMaMi28's antenna selection, and study whether it is affected by different \ac{UE} antennas. We start by the \ac{UE} rotation test in the static environment described in \fig\ref{fig:static_env}. A fixed spot, marked as ``A'' in \fig\ref{fig:static_env}, was selected for the rotation test. The distance between the spot ``A'' and the LuMaMi28 \ac{BS} was \SI{3}{\m}. Assuming that the direction to the \ac{BS} is $0^\circ$, two \acp{UE} co-located on the moving cart kept rotating between $-90^\circ$ and $90^\circ$, during the measurements. Real-time \ac{UL} throughputs of each \ac{UE} were simultaneously recorded while four consecutive experiments were separately performed for with\hspace{0.2em}/\hspace{0.2em}without antenna selection and different equalizers, \ie, \ac{MRC} and \ac{ZF}.

\fig\ref{fig:15_rotation} plots \acp{CDF} of the \acp{UE}' \ac{UL} throughput of LuMaMi28 with\hspace{0.2em}/\hspace{0.2em}without antenna selection. For \ac{MRC} equalizer, the median throughput-gain of \ac{UE}1 through antenna selection is higher than that of \ac{UE}2, \ie, $2.28\times$ for yagi and $1.12\times$ for patch antenna, as shown in \fig\ref{fig:15_mrc}. The \ac{UE}2 \ac{UL} throughput is, however, higher than that of \ac{UE}1, both with and without antenna selection. The \ac{UE}2 has a median throughput of $3.89$ and $3.48\hspace{0.2em}\rm{Mbit/s}$ while the \ac{UE}1 only achieves $0.66$ and $0.29\hspace{0.2em}\rm{Mbit/s}$, with and without antenna selection, respectively. It can be seen that two \acp{UE}' \ac{UL} signals interfere with each other, but \ac{UE}1 is affected by a larger \ac{IUI} from \ac{UE}2 since \ac{UE}2 has higher \ac{Tx} antenna gain. For \ac{ZF}, the \ac{CDF} curves of \ac{UE}1 and \ac{UE}2 are almost identical as shown in \fig\ref{fig:15_zf}. The \ac{UE}1 obtains a bit higher median throughput-gain than \ac{UE}2 by using antenna selection, \ie, $1.26\times$ for \ac{UE}1 and $1.21\times$ for \ac{UE}2. This result shows that, in case where the \ac{IUI} is eliminated by the \ac{ZF} equalizer, a larger beam-width of yagi antenna is useful for antenna selection in the rotating environment at the fixed spot. It translates into higher throughput gain than the use of patch antenna array.
\begin{figure*}[t!]
\centering
\subfigure[ ]
{
\includegraphics[width=3.39in]{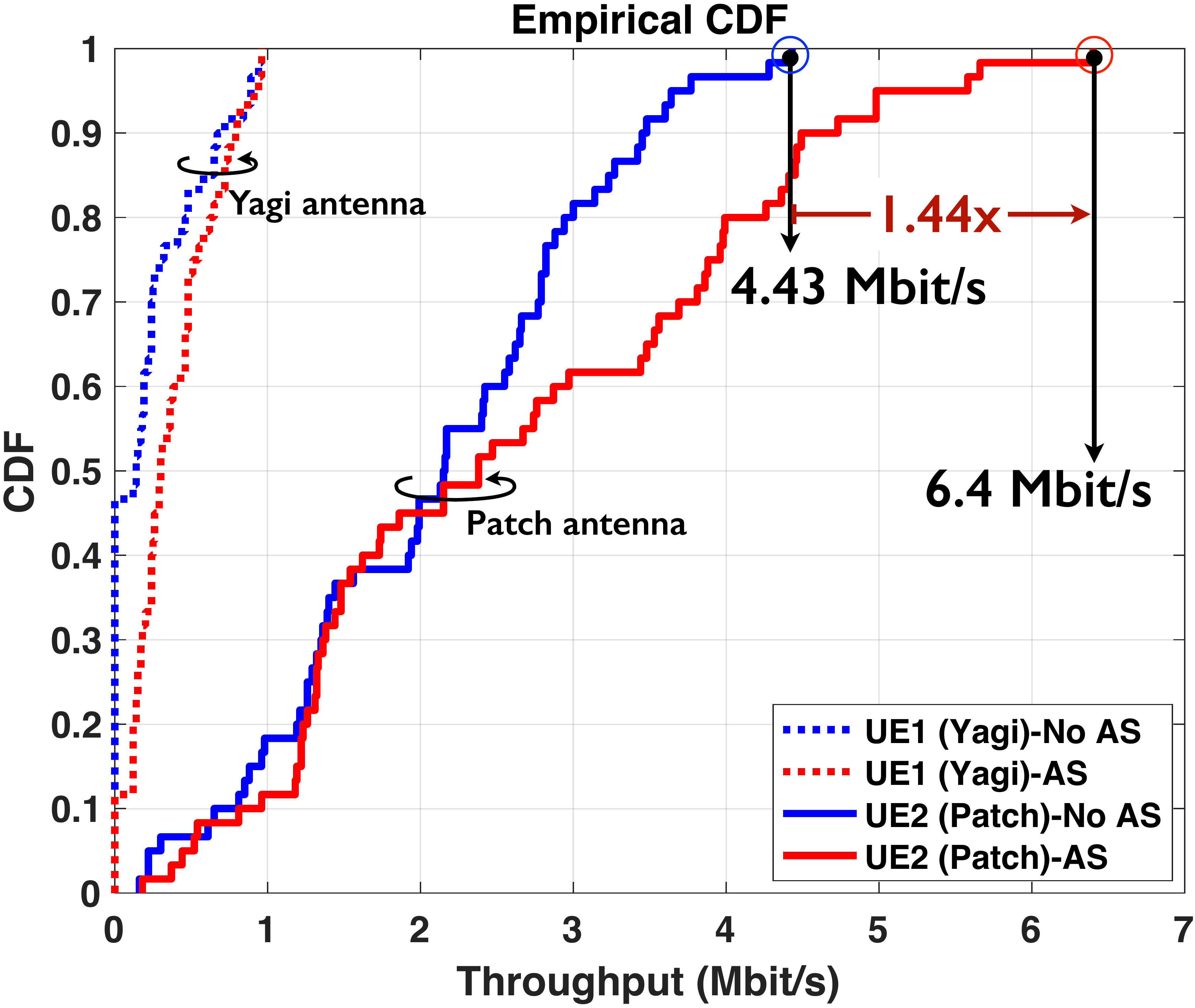}
\label{fig:17_mrc}
}
\subfigure[ ]
{
\includegraphics[width=3.39in]{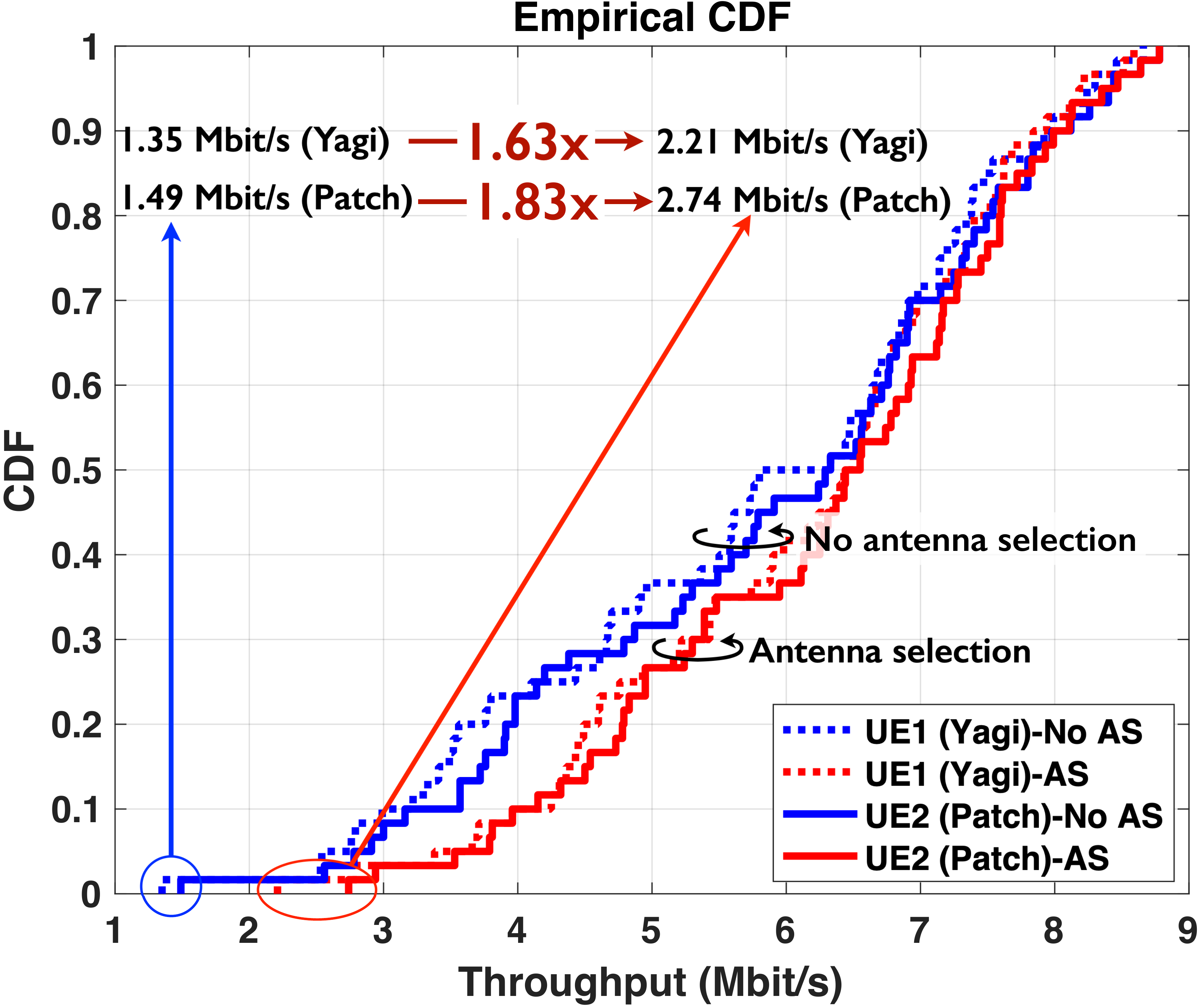}
\label{fig:17_zf}
}
\caption{Throughput \acp{CDF} of \acp{UE} moving in the vertical route~(around $3\hspace{0.2em}\rm{km/h}$). The dotted\hspace{0.2em}/\hspace{0.2em}solid lines indicate the \ac{CDF} of the \ac{UE}1 with yagi antenna and the \ac{UE}2 with patch antenna, respectively. The red\hspace{0.2em}/\hspace{0.2em}blue curves are for antenna selection\hspace{0.2em}/\hspace{0.2em}no antenna selection, respectively. (a) MRC (b) ZF.}
\label{fig:17_vertical}
\end{figure*}

\subsection{\ac{UE} Mobility Test} 
\label{subsec:mobility_test}
To evaluate the LuMaMi28's performance in mobility environments, we considered horizontal, vertical, and circled routes for mobility environments, as seen in \fig\ref{fig:mobility_env}.
\begin{figure*}[t!]
\centering
\subfigure[ ]
{
\includegraphics[width=3.39in]{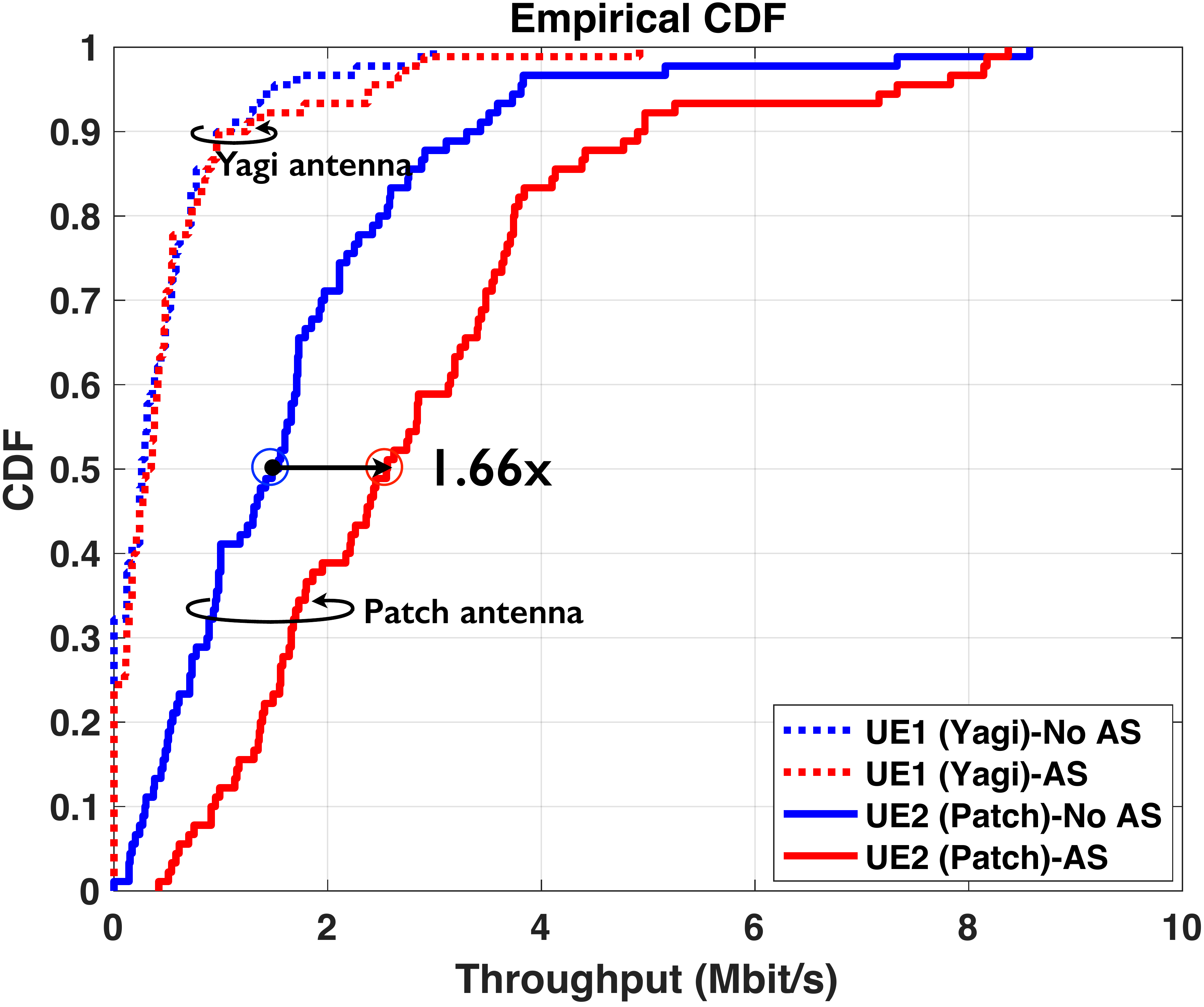}
\label{fig:18_mrc}
}
\subfigure[ ]
{
\includegraphics[width=3.39in]{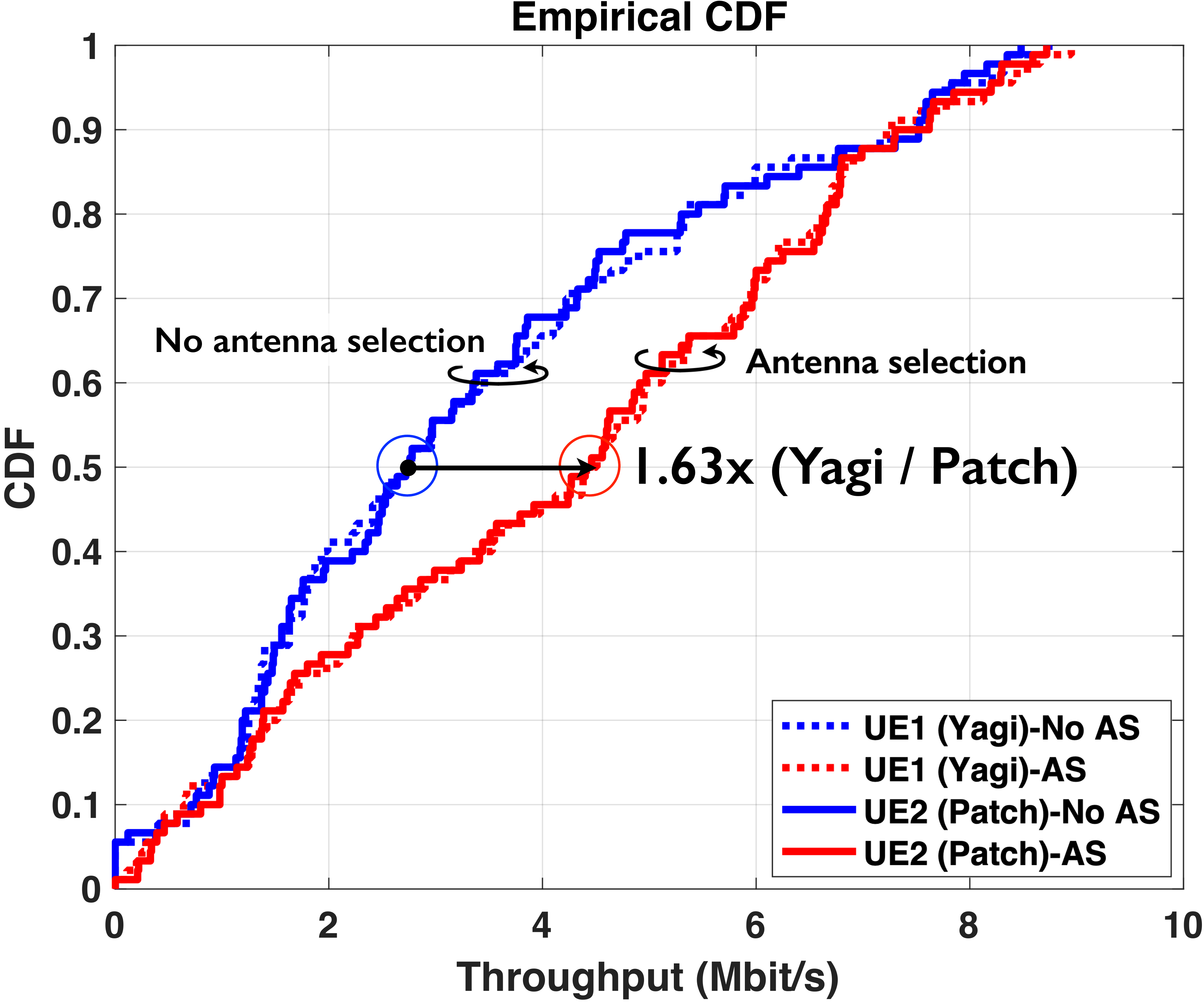}
\label{fig:18_zf}
}
\caption{Throughput \acp{CDF} of \acp{UE} moving in the circled route~(around $3\hspace{0.2em}\rm{km/h}$). The dotted\hspace{0.2em}/\hspace{0.2em}solid lines indicate the \ac{CDF} of the \ac{UE}1 with yagi antenna and the \ac{UE}2 with patch antenna, respectively. The red\hspace{0.2em}/\hspace{0.2em}blue curves are for antenna selection\hspace{0.2em}/\hspace{0.2em}no antenna selection, respectively. (a) MRC (b) ZF.} 
\label{fig:18_circled}
\end{figure*}

\subsubsection{Horizontal Route}
\fig\ref{fig:16_horizontal} plots \acp{CDF} of the \acp{UE}' \ac{UL} throughput in the horizontal route. With \ac{MRC}, the \ac{UE}1~(yagi) and \ac{UE}2~(patch) achieve a median throughput-gain of $3.32\times$ and $1.12\times$, respectively, through antenna selection. However, the \ac{UE}2 has $4\times$\hspace{0.1em}(antenna selection) and $6.54\times$\hspace{0.1em}(no antenna selection) higher median-throughput than the \ac{UE}1. When \ac{ZF} is applied, the difference between \ac{UE}1 and \ac{UE}2 is marginal. The throughput gain through antenna selection is $1.13\times$ and $1.15\times$ for the \ac{UE}1 and \ac{UE}2, respectively. For both \ac{MRC} and \ac{ZF}, a similar tendency is also seen, as compared to the \ac{UE} rotation test. The reason that the throughputs in this mobility test are generally worse, as compared to those in the rotation test, was affected by larger path loss and mobility environment.

\subsubsection{Vertical Route}
\fig\ref{fig:17_vertical} plots \acp{CDF} of the \acp{UE}' \ac{UL} throughput in the vertical route. For \ac{MRC}, the median throughput-gains of \ac{UE}1~(yagi) and \ac{UE}2~(patch), by using antenna selection, are marginal, as shown in \fig\ref{fig:17_mrc}. It can be seen that antenna selection is generally not very beneficial when moving to the \ac{BS} direction or its opposite. One interesting observation is that, there is a gap between the peak throughputs of \ac{UE}1 with\hspace{0.2em}/\hspace{0.2em}without antenna selection, \ie, $6.4\hspace{0.2em}\rm{Mbit/s}$ and $4.43\hspace{0.2em}\rm{Mbit/s}$, respectively, which is a $1.14\times$ difference. The difference of peak throughput could be related with narrower beam-width but higher gain of patch \ac{UE} antenna. It could provide a room for antenna selection, as the distance between from the LuMaMi28 \ac{BS} is close. For \ac{ZF}, the throughput curves between two \acp{UE} are almost same. Also, those between with\hspace{0.2em}/\hspace{0.2em}without antenna selection have no significant difference. It can be seen because the interference elimination by using \ac{ZF} dominantly affects each \ac{UL} throughput improvement. One remarkable thing is that, the antenna selection yields an improvement of lowest throughput of both \acp{UE} for \ac{ZF}, \ie, $1.63{\times}$~(\ac{UE}1) and $1.83{\times}$~(\ac{UE}2), as shown in \fig\ref{fig:17_zf}. It shows that the throughput gain by using antenna selection could be obtained in noise-limited situations, where the performance is not dominated by the \ac{ZF} equalizer. 

\subsubsection{Circled Route}
\fig\ref{fig:18_circled} plots \acp{CDF} of the \acp{UE}' \ac{UL} throughput in the circled route. For \ac{MRC}, \ac{UE}2~(patch) obtains a median throughput gain of $1.66\times$ through antenna selection while that of \ac{UE}1~(yagi) is relatively marginal. For \ac{ZF}, as with the previous experiments, there is no significant difference between the throughputs of two \acp{UE}. Also, by using the antenna selection, the two \acp{UE} achieve the same median throughput-gains ($1.63\times$). In both \ac{MRC} and \ac{ZF}, the two \acp{UE}' throughput values are placed over the larger range, as compared to the previous experiments.

\section{Concluding Remark}
\label{sec:con}
In this paper, we have presented LuMaMi28, real-time \SI{28}{\GHz} massive \ac{MIMO} systems with \ac{UE} beam steering. Furthermore, real-time measurement results of LuMaMi28 in static and mobility environments have been provided. To explore the impact of different \ac{UE} antennas, we adopted two kinds of array antennas: an yagi array antenna with wider angle coverage and a patch array antenna with higher gain. The main findings from our experiments are summarized as follows.
\begin{itemize}
\item The result of captured real-channel data shows that the achievable \ac{UE}-antenna selection gains are $10.0\hspace{0.2em}{\rm{dB}}$ and $12.8\hspace{0.2em}{\rm{dB}}$ for yagi and patch array antennas, respectively, in a distance range of $3 - \SI{11}{\m}$ between \ac{BS} and \ac{UE}.
\item The \ac{UE}-antenna selection of LuMaMi28 delivers a $3.32\hspace{0.1em}\times$\hspace{0.1em}(yagi)\hspace{0.2em}/\hspace{0.2em}$1.97\hspace{0.1em}\times$\hspace{0.1em}(patch) increase in the median \ac{UL} throughput in mobility environments, as compared to no antenna selection.
\item In case where the \ac{MRC} equalizer is employed at the \ac{BS}, it is found by overall measurements that the throughput of UE, equipped with patch array antenna, is higher than the \ac{UE} throughput with yagi array antenna, regardless of antenna selection. In case where the \ac{ZF} equalizer employed at the \ac{BS}, on the other hand, two \acp{UE}' throughputs have a similar tendency, resulting in almost the same throughput gain by using antenna selection.
\end{itemize}

We believe that these findings will be an important step forward in our understanding of the potential of \ac{UE} beam steering in \ac{mmWave} massive \ac{MIMO} systems. LuMaMi28 has been designed as a flexible testbed because it is implemented by a combination of \ac{FPGA}-based \acp{SDR} and extensible analog modules. The development of larger-scale massive \ac{MIMO} systems and novel baseband algorithms is an open area for further research with LuMaMi28.

\section*{Acknowledgment}
The authors would like to thank Dr. Carla D. Paola for providing Yagi array prototype foe test.

\ifCLASSOPTIONcaptionsoff
  \newpage
\fi

\renewcommand{\baselinestretch}{1.2}
\bibliographystyle{IEEEtran}
\bibliography{tmtt_21.bib}  
\input{tmtt_21.acro}

\end{document}